\documentclass[
    aps,
    prl,
    twocolumn,
    superscriptaddress,
    secnumarabic,
    floatfix,
]{revtex4-2}

\usepackage{psfrag}
\usepackage{multirow}
\usepackage{changepage}

\usepackage{booktabs}

\usepackage{algorithm}
\usepackage{algpseudocode}

\usepackage{CJKutf8}

\usepackage{url}
\usepackage{tikz}
\usepackage[symbol]{footmisc}
\usepackage{geometry}
\usepackage{setspace}
\usepackage{graphicx}
\usepackage{scalerel}

\usepackage{changepage}
\usepackage[american]{babel}
\usepackage{mathrsfs,amsmath}
\usepackage{wrapfig}
\usepackage{cases, mathtools}
\usepackage{amssymb, amsbsy} 
\usepackage{mathrsfs,amsmath}
\usepackage{listings}
\usepackage{bm}

\usepackage[colorlinks]{hyperref}
\usepackage{xcolor}
\hypersetup{
    colorlinks,
    linkcolor={red},
    citecolor={red},
    urlcolor={red}
}

\definecolor{lightblue}{RGB}{6,69,173}
\definecolor{myRED}{RGB}{149, 37, 51}
\definecolor{myBLUE}{RGB}{45, 115, 184}

\usepackage[colorlinks]{hyperref}
\usepackage{xcolor}
\hypersetup{
    colorlinks,
    linkcolor={lightblue},
    citecolor={lightblue},
    urlcolor={lightblue}
}

\usetikzlibrary{shapes.geometric}
\usepackage{rotating}
\usetikzlibrary{quotes,angles}
\usetikzlibrary{patterns}
\usetikzlibrary{decorations.pathmorphing}
\usetikzlibrary{decorations.markings}
\usetikzlibrary{matrix}
\usetikzlibrary{angles, quotes}
\usetikzlibrary{calc}

\renewcommand\thesection{\arabic{section}}

\def\XXint#1#2#3{{\setbox0=\hbox{$#1{#2#3}{\int}$}
     \vcenter{\hbox{$#2#3$}}\kern-.5\wd0}}

\usepackage{listings}
\usepackage{color}
 
\definecolor{codegreen}{rgb}{0,0.6,0}
\definecolor{codegray}{rgb}{0.5,0.5,0.5}
\definecolor{codepurple}{rgb}{0.58,0,0.82}
\definecolor{backcolour}{rgb}{0.95,0.95,0.92}
 
\lstdefinestyle{mystyle}{
    backgroundcolor=\color{backcolour},   
    commentstyle=\color{codegreen},
    keywordstyle=\color{magenta},
    numberstyle=\tiny\color{codegray},
    stringstyle=\color{codepurple},
    basicstyle=\footnotesize,
    breakatwhitespace=false,         
    breaklines=true,                 
    captionpos=b,                    
    keepspaces=true,                 
    numbers=left,                    
    numbersep=5pt,                  
    showspaces=false,                
    showstringspaces=false,
    showtabs=false,                  
    tabsize=2
}
 
\definecolor{myGreen}{RGB}{0,128,0}
\definecolor{myPurple}{RGB}{139,0,139}
\definecolor{myRed}{RGB}{162, 20, 47}
\definecolor{myYellow}{RGB}{255, 215, 0}
\definecolor{myGrey}{RGB}{189, 201, 199}

\usepackage{float,yfonts}

\newcommand{\ii}{\mathrm{i}}
\newcommand{\ee}{\mathrm{e}}
\newcommand{\dd}{\mathrm{d}}

\begin{document}

\setcounter{secnumdepth}{3}

\title{Self-Referenced Ultrafast Acousto-Optics Resolves Strain, Refractive Index, and Photoelasticity}

\author{Richard Wiltshaw}
\email{r.wiltshaw17@imperial.ac.uk}
\affiliation{Graduate School of Engineering, The University of Osaka, Suita, Osaka 565-0871, Japan}
\affiliation{Department of Mathematics, Imperial College London, London SW7 2AZ, United Kingdom}

\author{Motonobu Tomoda}
\affiliation{Faculty of Science and Technology, Oita University, Oita 870-1192, Japan}

\author{Osamu Matsuda}
\affiliation{Division of Applied Physics, Graduate School of Engineering, Hokkaido University, Sapporo 060-8628, Japan}

\author{Takehiro Tachizaki}
\affiliation{Institute for Chemical Research, Kyoto University, Uji, Kyoto 611-0011, Japan}

\author{Hirotsugu Ogi}
\affiliation{Graduate School of Engineering, The University of Osaka, Suita, Osaka 565-0871, Japan}

\author{Oliver B. Wright}
\email{olly@eng.hokudai.ac.jp}
\affiliation{Graduate School of Engineering, The University of Osaka, Suita, Osaka 565-0871, Japan}
\affiliation{Hokkaido University, Sapporo 060-0808, Japan}

\begin{abstract}
When an unknown physical field is observed through unknown material properties, determining either requires knowledge of the other. We show that one normal-incidence complex optical reflectance transient from a picosecond acoustic pulse yields the complex refractive index, photoelastic constant, and strain profile simultaneously. Surface displacement provides an internal reference for photoelastic scattering. In chromium, the recovered refractive index agrees with independent ellipsometry, while the strain follows a two-temperature-model prediction. The approach offers simultaneous two-dimensional imaging of refractive index, photoelasticity, and strain, and suggests a general self-referencing principle for other dynamical fields.
\end{abstract}


\maketitle

Optical measurements can reveal otherwise inaccessible structure and material properties through their effects on transmitted, reflected, or scattered light. Tomographic and interferometric methods obtain additional information by varying incidence angle, optical frequency, or path delay \cite{Fercher2003OCT,Choi2007TomographicPhase,Fercher1995SpectralInterferometry}. A more demanding situation arises when the measured light depends on two quantities that are both unknown. Some methods exploit redundancy across many measurements to determine the two unknowns together. Ptychographic diffractive imaging, for example, can recover an object and the illuminating wavefront from overlapping diffraction measurements \cite{Thibault2008,Maiden2009}, whereas coherent Raman spectroscopy can retrieve the molecular vibrational response and optical probe envelope from a two-dimensional delay--frequency measurement \cite{Li2026BlindFROG}. When probing elastic fields, this problem takes a particularly direct form because strain changes the optical properties of a material through photoelasticity. Photoelastic methods have long been used to visualize acoustic fields \cite{Dally1980DynamicPhotoelasticity,Liu2020UltrasonicVisualization,Lu2022TransientShear,Saito2010Photoelastic}, and photoelastic coefficients can be determined by applying known external stresses \cite{Brito2005SelfCalibrating}. Without such external information, as in the above examples with two unknowns, recovering the elastic field requires knowledge of the photoelastic properties, whereas determining those properties from the same optical measurement requires knowledge of the field.

This circular dependence is particularly acute in picosecond ultrasonics,
where GHz--THz strain pulses are generated in solids by ultrashort optical
pump pulses and their reflections detected as acoustic echoes by probe pulses
\cite{Thomsen1986Surface,Matsuda2015Fundamentals}. For an isotropic opaque
solid with longitudinal acoustic propagation and optical probing normal to the
surface, the detected probe field is a coherent sum of surface reflection and
light scattered by strain at different depths
\cite{Matsuda2002Reflection,Matsuda2015Fundamentals}. The dielectric
constant $\widetilde{\epsilon}=\widetilde{n}^{\,2}$, where
$\widetilde n$ is the complex refractive index, governs surface reflection,
optical phase accumulation, and attenuation with depth, whereas
$\widetilde P_{12}$ governs local strain-induced changes in
$\widetilde{\epsilon}$. The strain profile sets the amplitude, sign, and depth
of the scattering contributions, and free-surface motion adds an optical phase
change. The measured reflectance amplitude and phase therefore arise from
depth-integrated coherent interference in which the dielectric properties, photoelasticity, and strain field are strongly intertwined.

Earlier ultrafast optical approaches have disentangled subsets of these
quantities using additional measurements or demanding geometries. In transparent materials, time-domain Brillouin tomography can
recover refractive index and sound velocity by scanning the optical incidence
angle \cite{Tomoda2023AngleScan}, and the strain profile when the refractive
index is known \cite{Tomoda2023Tomographic}. For opaque films, ultrafast
ellipsometric interferometry can isolate the strain profile without prior
knowledge of the dielectric or photoelastic constants by cancelling the photoelastic contribution to the acoustic echo using combined $s$- and $p$-polarized responses at oblique incidence \cite{Matsuda2013Ellipsometric}. More recently,
normal-incidence measurements of picosecond echo amplitude and phase have
recovered the strain profile with $\widetilde n$ and $\widetilde P_{12}$ known
\cite{Tachizaki2023SpectralAnalysis}, and then $\widetilde P_{12}$ together
with the strain profile with $\widetilde n$ known
\cite{tomoda2026simultaneous}. Thus $\widetilde n$ still requires an
independent optical measurement
\cite{Nilsson1968NormalIncidence,Schulz1951Interferometric,Fujiwara2007},
and simultaneous recovery of all three quantities has remained out of reach. 

Here we show that this mutual dependence can be broken by the physics of the optical measurement itself. The same strain field produces both free-surface motion, whose optical coupling is known, and light scattering governed by the unknown refractive index and photoelastic constant, so that the surface-motion contribution provides an internal reference. Because the different acoustic frequency components are governed by the same material properties and the strain field must be real, a single complex reflectance transient contains enough information to resolve the complex refractive index, photoelastic constant, and strain profile without angle, wavelength, or polarization scans or independently supplied optical or photoelastic constants. More generally, the same self-referencing principle applies whenever a time-dependent field produces surface motion as well as a change in optical properties.

We consider a plane-wave optical probe at normal incidence on the free
surface $z=0$ of an isotropic opaque film. During normal acoustic reflection at the surface, we treat the strain field $\eta(z,t)$ as a
one-dimensional longitudinal plane wave of velocity $\nu$, neglecting attenuation and dispersion. The film thickness exceeds both the
strain pulse width and the probe optical penetration depth.

The condition of zero normal stress
at the free surface implies strain reversal on reflection, giving
$\eta(z,t)=\eta_A(t+z/\nu)-\eta_A(t-z/\nu)$, where $\eta_A(t)$ is the 
incident strain waveform. The incident strain waveform $\eta_A(t)$ and its spectrum
$ A(\omega)$ are related by the Fourier transform pair
$ A(\omega)=(2\pi)^{-1}\int_{-\infty}^{\infty}
\eta_A(t)\ee^{\ii\omega t}\dd t$ and
$\eta_A(t)=\int_{-\infty}^{\infty}
 A(\omega)\ee^{-\ii\omega t}\dd\omega$.
Here $\omega$ is the acoustic angular frequency and, since
$\eta_A(t)$ is real,
$ A(\omega)= A^{*}(-\omega)$, where the
superscript $*$ denotes complex conjugation.

Let $ r$ denote the unperturbed complex optical reflectance and echo  
$\delta r(t)/ r=\rho(t)+i\delta\phi(t)$ its transient change, where $\rho(t)$
is the relative amplitude reflectance change and $\delta\phi(t)$ is the
optical phase change. We define $k_0 = 2\pi/\lambda$, where $\lambda$ is the probe vacuum
wavelength and
$\widetilde n=n+i\kappa$ (assuming $\kappa>0$). The relevant complex
photoelastic constant $\widetilde P_{12}$ relates the local dielectric
perturbation to the longitudinal strain through
$\Delta\widetilde\epsilon=\widetilde P_{12}\eta$.

The optical reflectance change couples $\widetilde n$ and
$\widetilde P_{12}$. Strained material at depth $z'$ produces a local
dielectric perturbation and scatters probe light. The contribution from each
depth $z'$ returns to the surface and interferes coherently with the surface
reflection, with propagation factor
$\ee^{2\ii\widetilde n k_0z'}=
\ee^{2\ii n k_0z'}\ee^{-2\kappa k_0z'}$ accounting for round-trip phase
and optical attenuation. The measured field is therefore the coherent sum over the
strain distribution sampled by the probe. Including the phase change from
free-surface motion, to first order in strain,
\cite{Matsuda2002Reflection,Matsuda2015Fundamentals}
\begin{equation}
\frac{\delta r(t)}{ r}
=
\frac{2\ii k_0\widetilde P_{12}}
{1-\widetilde n^{2}}
\int_{0}^{\infty}
\eta(z',t)\,
\ee^{2\ii\widetilde n k_0z'}\,\dd z'
-
2\ii k_0u(0,t),
\label{eq:theory_reflectance_integral}
\end{equation}
where $u(0,t)=\int_{0}^{\infty}\eta(z',t)\dd z'$ is the displacement of the
free surface. Thus $\widetilde n$ controls optical propagation and attenuation,
$\widetilde P_{12}$ couples light to strain, and $\eta(z,t)$ sets the
distribution of scatterers. Equation~\eqref{eq:theory_reflectance_integral} makes the circular
dependence explicit: recovering $\eta(z,t)$ from the complex echo requires
$\widetilde n$ and $\widetilde P_{12}$, whereas determining
$\widetilde n$ and $\widetilde P_{12}$ from the echo requires
$\eta(z,t)$.

To remove offsets and varying backgrounds from $\delta r(t)$, we follow Refs.~\cite{Tachizaki2023SpectralAnalysis,
tomoda2026simultaneous} by differentiating the measured variation before taking its
temporal Fourier transform using the convention defined above: 
$ R(\omega)=\mathcal{F}[
\dd(\delta r/ r)/\dd t]$. Evaluating the depth integral gives
\begin{equation}
4\ii k_0\nu
\left[
1+
\frac{\omega^{2}\widetilde P_{12}}
{(\widetilde n^{2}-1)
(\omega^{2}-4\widetilde n^{2}k_0^{2}\nu^{2})}
\right]
 A(\omega)
=
 R(\omega).
\label{echo1}
\end{equation}
The measured echo determines $R(\omega)$, whereas $k_0$ and $\nu$ are known experimental quantities; $\widetilde n$, $\widetilde P_{12}$, and $A(\omega)$ are unknown. In Eq.~\eqref{echo1}, the term 1 arises from free-surface motion and is independent of $\widetilde n$ and $\widetilde P_{12}$, whereas the second term describes light scattering governed by these material properties. The surface-motion term therefore provides an internal reference for the material-dependent scattering. Crucially, the different acoustic frequency components are governed by the same $\widetilde n$ and $\widetilde P_{12}$, and the reality of the strain field requires $A(\omega)=A^{*}(-\omega)$. Together, these constraints provide enough information to determine $\widetilde n$, $\widetilde P_{12}$, and the strain spectrum from a single complex reflectance transient.

Let $G(\omega)$ denote the factor in square brackets in Eq.~\eqref{echo1}, and define the measured ratio $J(\omega)=R^{*}(-\omega)/R(\omega)$. Combining Eq.~\eqref{echo1} with its complex conjugate at $-\omega$ and using $A(\omega)=A^{*}(-\omega)$ gives, for each retained frequency with nonzero strain amplitude,
\begin{equation}
J(\omega)G(\omega)+G^{*}(\omega)=0.
\label{materialconstraint}
\end{equation}
Thus the unknown strain amplitude has been eliminated, and every retained frequency provides a constraint on the same $\widetilde n$ and $\widetilde P_{12}$. We determine the material constants by requiring Eq.~\eqref{materialconstraint} to be satisfied across the retained frequencies, without assuming a strain profile; Eq.~\eqref{echo1} then gives the corresponding strain spectrum directly. The retained-frequency equations can be cast as a multiparameter nonlinear
eigenvalue problem; details of its numerical solution are given in
Appendix~\ref{APPEN} and the Supplemental Material.

We now apply this approach to experimental picosecond ultrasonic echoes from a
$190$~nm polycrystalline Cr film on a Si(100) substrate, with
$\nu=6650$~m~s$^{-1}$
\cite{saito2003picosecond,saito2004erratum}. Pump light at
$\lambda_p=415$~nm generates $\eta(z,t)$ through thermoelastic stress. The pulse reflects at the Cr/Si interface and returns to the vicinity of the Cr
surface, where the strain scatters probe light at $\lambda=830$~nm, focused at
normal incidence to an approximately $20~\mu\mathrm{m}$ full-width at
half-maximum spot, whereas the surface motion changes its phase
\cite{saito2003picosecond,saito2004erratum}.

\begin{figure}[H]
\vspace*{-0.2cm}
\hspace*{+0.5cm}
\begin{tikzpicture}[x=1bp,y=-1bp]
  \path[use as bounding box] (-28.20393,-5.59970) rectangle (179.86254,203.91239);
  \begin{scope}[overlay,shift={(-65.4471299,-16.5589124)},scale=0.525679758,transform shape]
    \clip (71.7500,21.7500) rectangle (465.7500,418.5000);
    \node[anchor=north west,inner sep=0pt,outer sep=0pt] at (0,0)
      {\includegraphics[width=560bp]{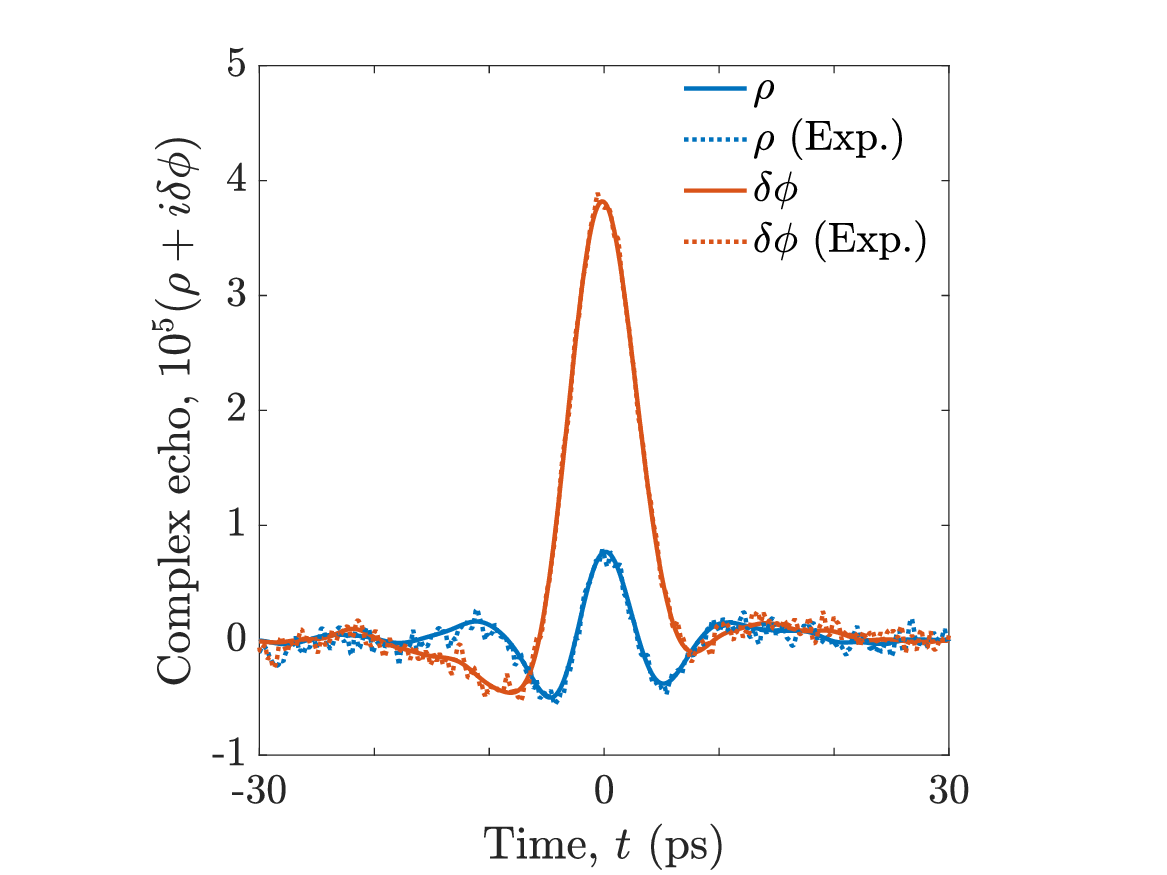}};
    \fill[white,overlay] (283.18302,409.20780) rectangle (286.55502,416.59609);
    \fill[white,overlay] (88.14140,190.36999) rectangle (95.52931,193.74198);
  \end{scope}
\end{tikzpicture}
\vspace*{-0.2cm}
\caption{Experimental (Exp.) complex relative reflectance change
$\delta r/ r=\rho+i\delta\phi$ for Cr
\cite{saito2003picosecond,saito2004erratum}, compared with the complex
relative reflectance change calculated from the recovered material constants
and strain field.}
\label{fig:BSwithReflectanceCr}
\end{figure}

Figure~\ref{fig:BSwithReflectanceCr} compares the measured $\rho(t)$ and
$\delta\phi(t)$ with values calculated from the recovered material constants
and strain field.

\begin{figure}[H]
\vspace*{-0.0cm}
\hspace*{0.2cm}
\begin{tikzpicture}[x=1bp,y=-1bp,scale=1.1, transform shape]
  \path[use as bounding box] (-46.60094,-1.11906) rectangle (164.35375,341.16156);
  \begin{scope}[overlay,shift={(-60.0075000,-15.0018750)},scale=0.476250000,transform shape]
    \clip (29.2500,30.2500) rectangle (470.0000,359.7500);
    \node[anchor=north west,inner sep=0pt,outer sep=0pt] at (0,0)
      {\includegraphics[width=560bp]{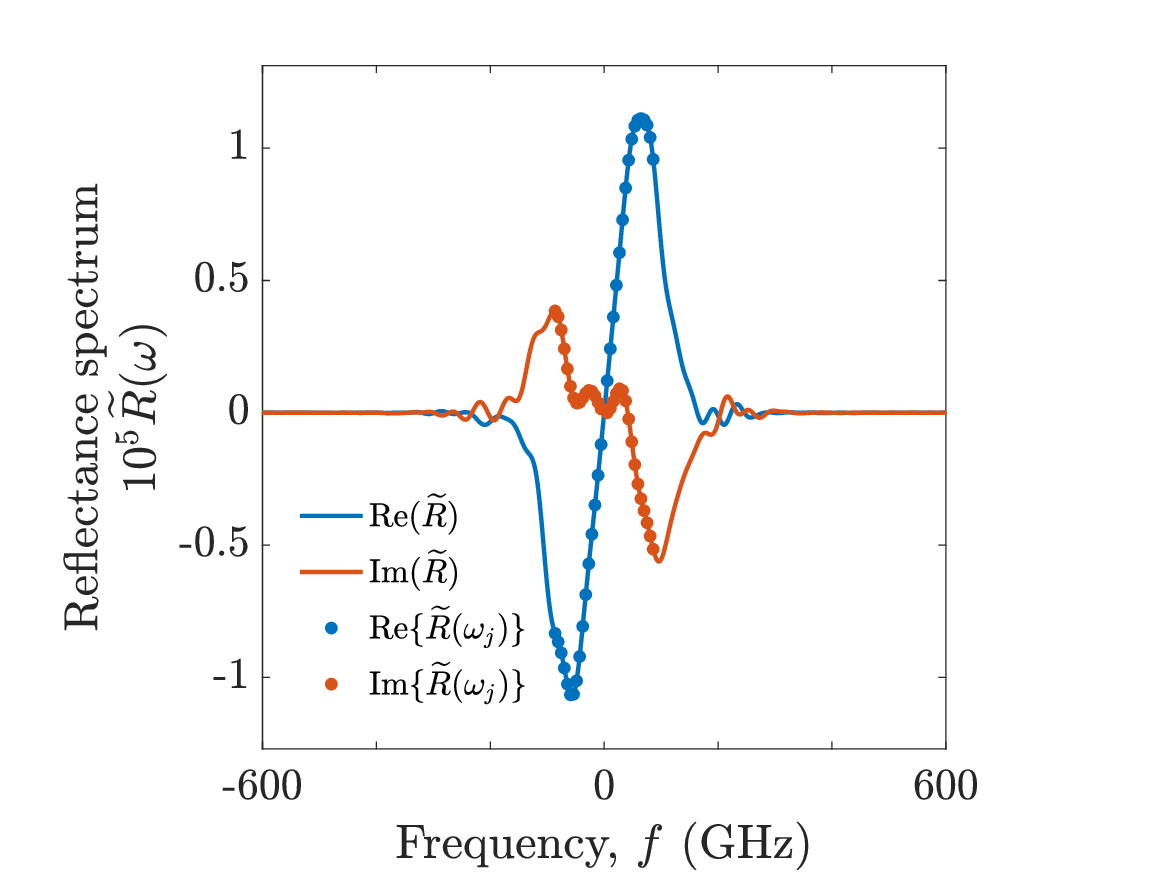}};
    \fill[white,overlay] (51.65159912,187.34299011) rectangle (54.93950348,200.70901184);
    \fill[white,overlay] (204.90300293,236.62900085) rectangle (214.39399414,239.10400085);
    \fill[white,overlay] (205.12500305,263.55198975) rectangle (214.61599426,266.02698975);
    \fill[white,overlay] (206.68001099,290.47398682) rectangle (216.17100220,292.94898682);
    \fill[white,overlay] (206.90300293,317.39598389) rectangle (216.39399414,319.87098389);
  \end{scope}
  \begin{scope}[overlay,shift={(-60.0075000,141.2081250)},scale=0.476250000,transform shape]
    \clip (35.5000,31.2500) rectangle (470.0000,418.7500);
    \node[anchor=north west,inner sep=0pt,outer sep=0pt] at (0,0)
      {\includegraphics[width=560bp]{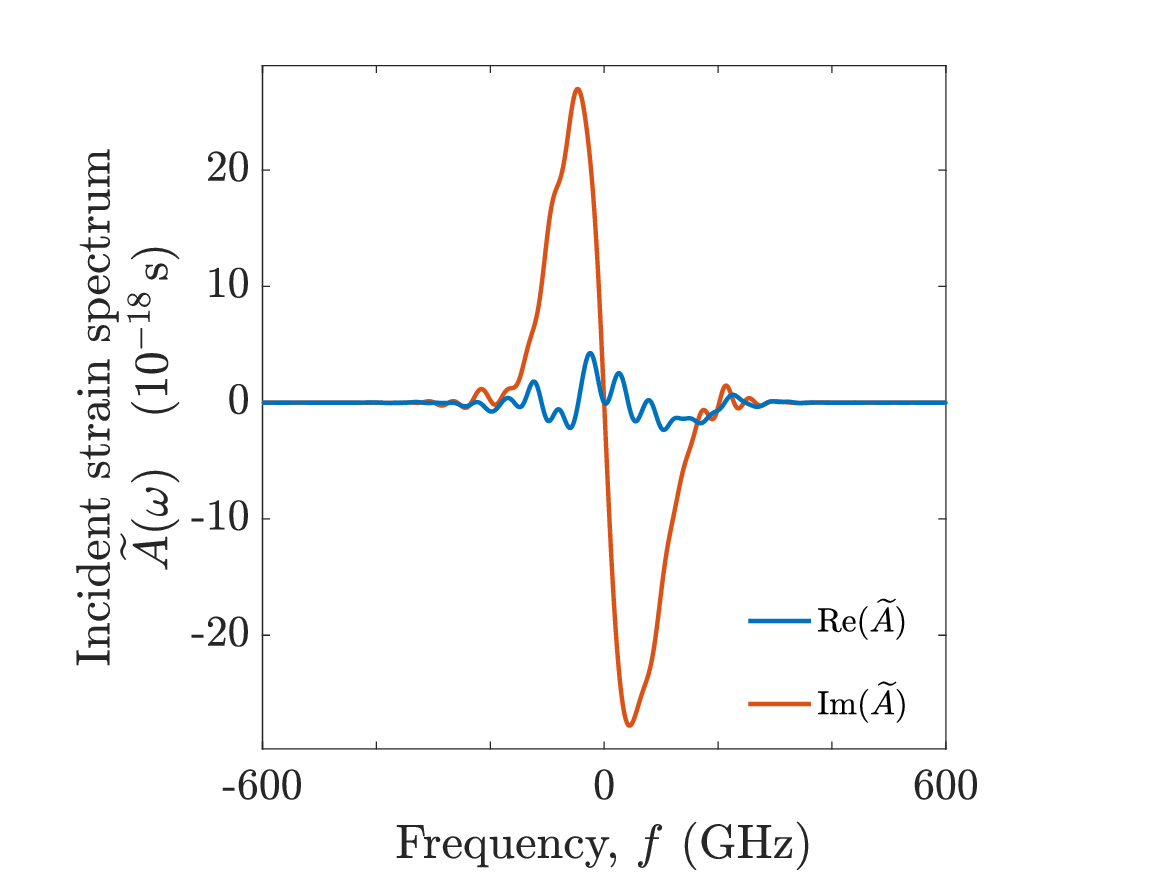}};
    \fill[white,overlay] (57.48480148,255.74498291) rectangle (60.77299957,269.10999756);
    \fill[white,overlay] (420.80699463,287.10698242) rectangle (430.29801636,289.58198242);
    \fill[white,overlay] (421.02901001,326.91948242) rectangle (430.52003174,329.39448242);
    \fill[white,overlay] (292.50202,408.92301) rectangle (295.98303,416.60702);
  \end{scope}
  \node[overlay,anchor=north west,inner sep=0pt,outer sep=0pt,
        font=\fontsize{11}{12}\selectfont] at (5.0000,5.0000) {$(a)$};
  \node[overlay,anchor=north west,inner sep=0pt,outer sep=0pt,
        font=\fontsize{11}{12}\selectfont] at (5.0000,161.2100) {$(b)$};
\end{tikzpicture}

\vspace*{-0.25cm}
\caption{Spectral recovery from the experimental Cr echo.
(a) Measured $R(f)$; the dots mark the $32$ frequencies used to determine the material constants from Eq.~\eqref{materialconstraint}. (b) Strain spectrum for the selected solution.}
\label{fig:RaAaEtaACr}
\end{figure}

Differentiating and Fourier transforming the measured echo gives
$R(\omega)$, plotted against $f=\omega/(2\pi)$ in
Fig.~\ref{fig:RaAaEtaACr}(a). We determine possible values of
$\widetilde n$ and $\widetilde P_{12}$ from
Eq.~\eqref{materialconstraint} at the $32$ marked frequencies, without
assuming a strain profile. Equation~\eqref{echo1} then gives the corresponding
strain spectrum directly from the measured $R(\omega)$. Experimental uncertainty permits several such material-parameter pairs, which we
distinguish by requiring the reconstructed spectrum to satisfy the real-strain
condition $A(\omega)=A^{*}(-\omega)$ over the usable measured frequency range.
The selected solution gives the smallest normalized departure from this
condition. Details of the frequency selection and reconstruction are
given in the Supplemental Material.

For the selected solution, we obtain
$\widetilde n=3.27+2.85i$ and
$\widetilde P_{12}=35.0+4.06i$. The recovered $n$ and $\kappa$ agree with
independent ellipsometric values for the same Cr film
\cite{saito2003picosecond}, and the reconstructed complex echo agrees more
closely with the measurements than the model of Saito \textit{et al.}
\cite{saito2003picosecond,saito2004erratum}.

\begin{figure}[H]
\vspace*{-0.25cm}
\hspace*{-0.25cm}
\begin{tikzpicture}[x=1bp,y=-1bp]
  \path[use as bounding box] (-48.17798,-1.16340) rectangle (183.90013,207.66005);
  \begin{scope}[overlay,shift={(-42.9204406,-43.8463892)},scale=0.435740514,transform shape]
    \pgfseteorule
    \clip (11.7500,99.2500) rectangle (518.7500,305.1250)
                        (11.7500,120.0000) rectangle (68.0000,285.0000);
    \pgfsetnonzerorule
    \node[anchor=north west,inner sep=0pt,outer sep=0pt] at (0,0)
      {\includegraphics[width=560bp]{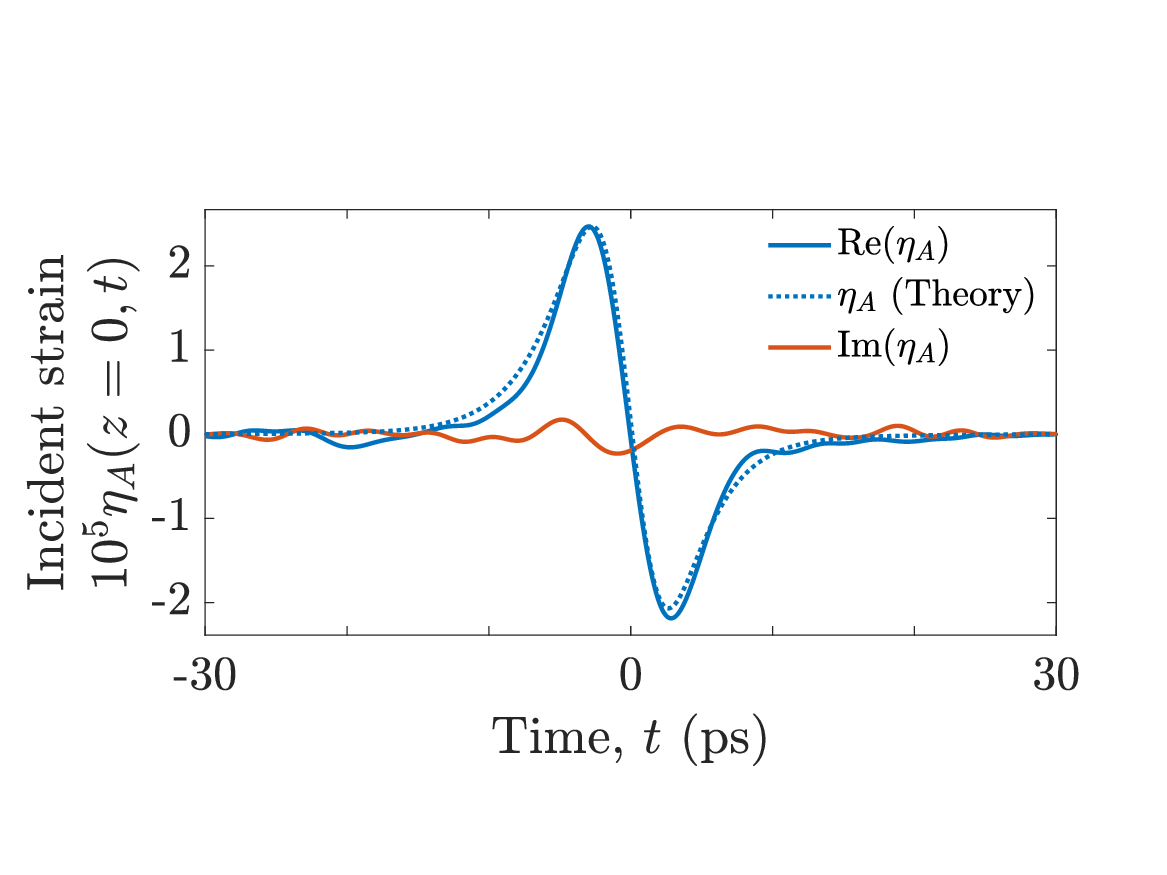}};
  \end{scope}
  \begin{scope}[overlay,shift={(-52.251789071,-43.8463892)},scale=0.435740514,transform shape]
    \clip (11.7500,120.0000) rectangle (68.0000,285.0000);
    \node[anchor=north west,inner sep=0pt,outer sep=0pt] at (0,0)
      {\includegraphics[width=560bp]{Figs/RconStrainCr.eps}};
  \end{scope}
  \begin{scope}[overlay,shift={(-53.1037132,41.0813060)},scale=0.455825864,transform shape]
    \clip (12.0000,103.0000) rectangle (518.7500,364.2500);
    \node[anchor=north west,inner sep=0pt,outer sep=0pt] at (0,0)
      {\includegraphics[width=560bp]{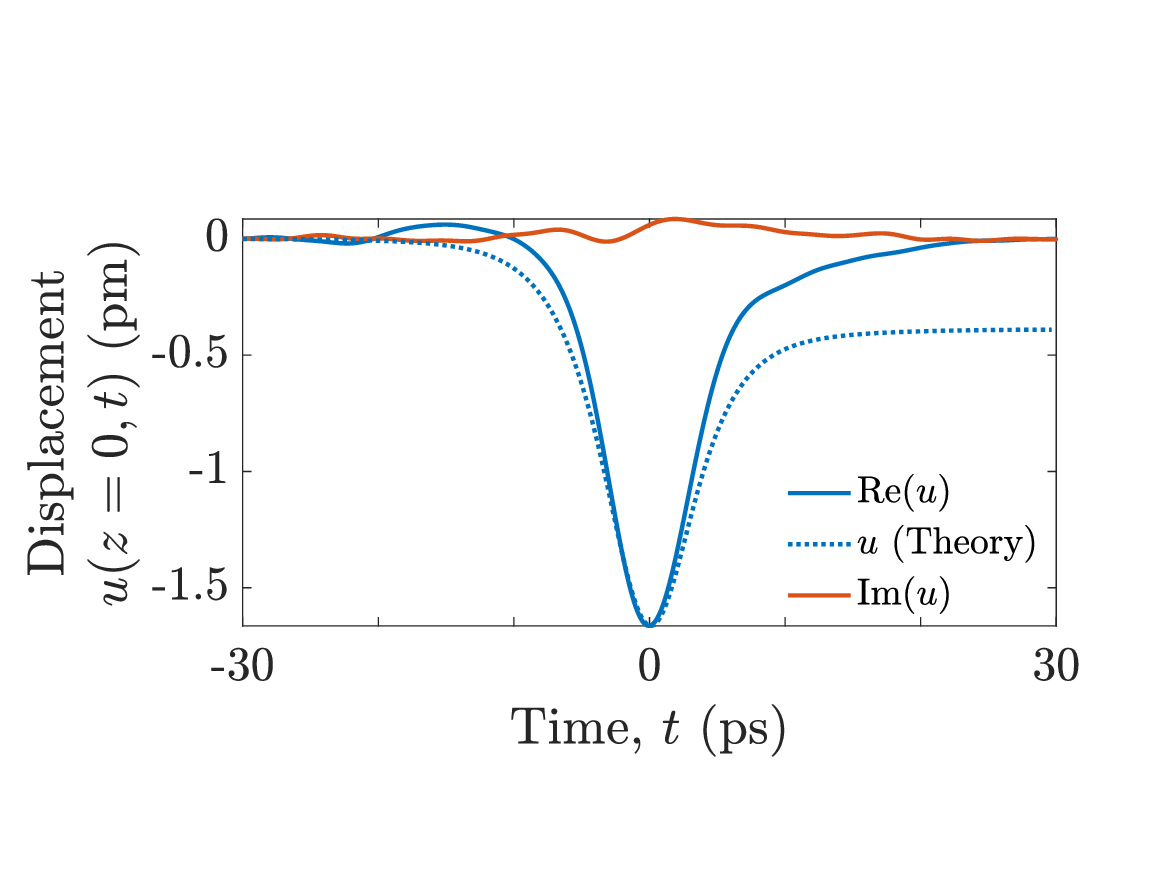}};
    \fill[white,overlay] (303.93601,353.72700) rectangle (307.66699,362.02111);
  \end{scope}
  \node[overlay,anchor=north west,inner sep=0pt,outer sep=0pt,
        font=\fontsize{11}{12}\selectfont] at (5.0000,5.0000) {$(a)$};
  \node[overlay,anchor=north west,inner sep=0pt,outer sep=0pt,
        font=\fontsize{11}{12}\selectfont] at (5.0000,160.0000) {$(b)$};
\end{tikzpicture}

\vspace*{-0.4cm}
\caption{Recovered time-domain quantities for the experimental Cr echo.
(a) Incident strain waveform obtained by inverse Fourier transformation of
the selected spectrum in Fig.~\ref{fig:RaAaEtaACr}(b).
(b) Corresponding free-surface displacement.}
\label{fig:CrTimeRecovery}
\end{figure}

Figure~\ref{fig:RaAaEtaACr}(b) shows the corresponding strain spectrum,
whose inverse Fourier transform gives the incident strain waveform in
Fig.~\ref{fig:CrTimeRecovery}(a) and the free-surface displacement in
Fig.~\ref{fig:CrTimeRecovery}(b). The recovered asymmetric strain closely follows the prediction (dashed curve)
based on the two-temperature model for electron and thermal diffusion, with
quadratic ultrasonic attenuation
$\alpha(f)=150f^2~\mathrm{m}^{-1}$ ($f$ in GHz) included during propagation
\cite{saito2003picosecond,saito2004erratum}. The corresponding predictions
without attenuation are shown in the Supplemental Material. Because the free-surface displacement is obtained by integrating the strain
over depth, small residual strain differences accumulate, so the agreement is
less close than for the strain.

Using the recovered material constants and strain field in
Eq.~\eqref{eq:theory_reflectance_integral} gives the calculated curves in
Fig.~\ref{fig:BSwithReflectanceCr}.

\begin{figure}[H]
\vspace*{-0.25cm}
\hspace*{-0.6cm}
\begin{tikzpicture}[x=1bp,y=-1bp,scale=1.1, transform shape]
  \path[use as bounding box] (-21.57143,-10.00215) rectangle (224.77953,89.05655);
  \begin{scope}[overlay,shift={(-29.4725275,-9.2307692)},scale=0.263736264,transform shape]
    \clip (32.7500,1.2500) rectangle (490.7500,299.2500);
    \node[anchor=north west,inner sep=0pt,outer sep=0pt] at (0,0)
      {\includegraphics[width=560bp]{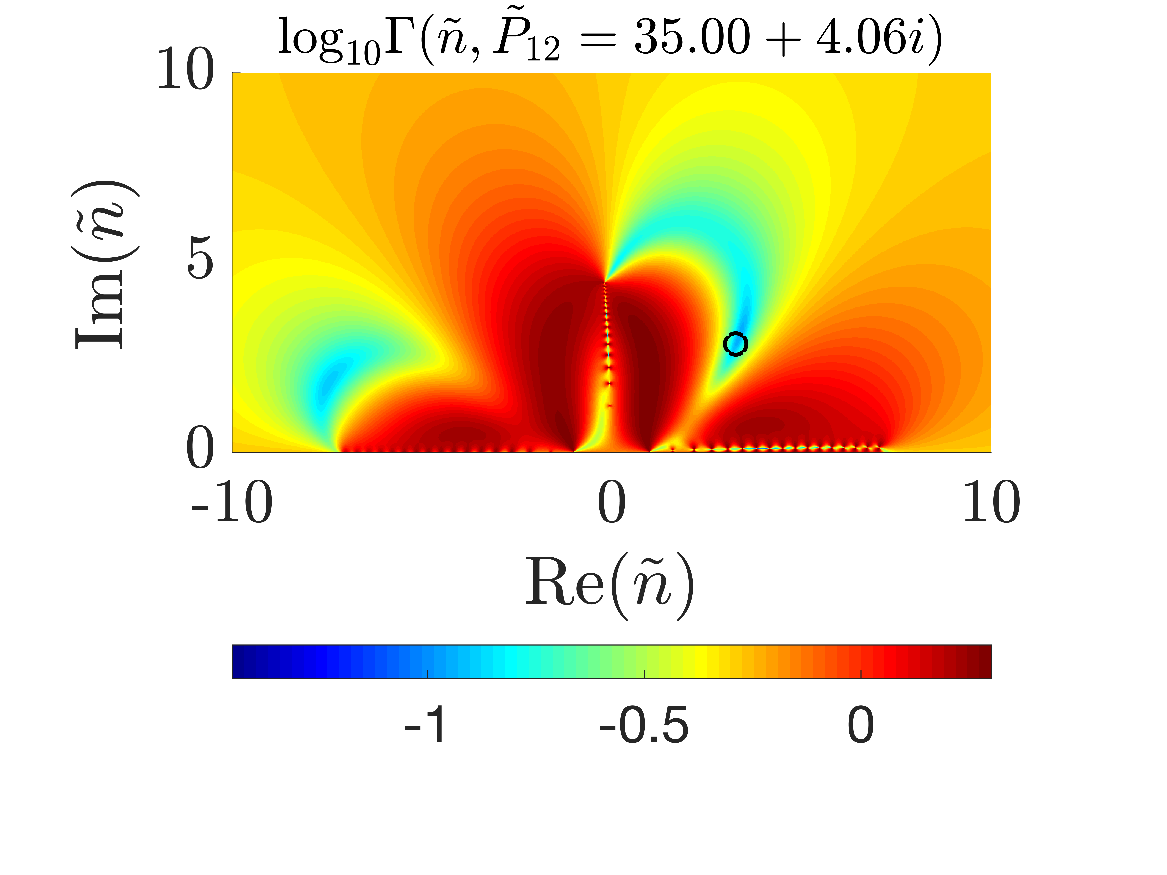}};
  \end{scope}
  \begin{scope}[overlay,shift={(-29.4725275,-6.5604396)},scale=0.263736264,transform shape]
    \clip (110.5000,308.2500) rectangle (477.0000,357.7500);
    \node[anchor=north west,inner sep=0pt,outer sep=0pt] at (0,0)
      {\includegraphics[width=560bp]{Figs/residualOverNExpCr.eps}};
  \end{scope}
  \begin{scope}[overlay,shift={(90.8775948,-9.8954903)},scale=0.274874732,transform shape]
    \clip (36.0000,2.2500) rectangle (484.2500,295.5000);
    \node[anchor=north west,inner sep=0pt,outer sep=0pt] at (0,0)
      {\includegraphics[width=560bp]{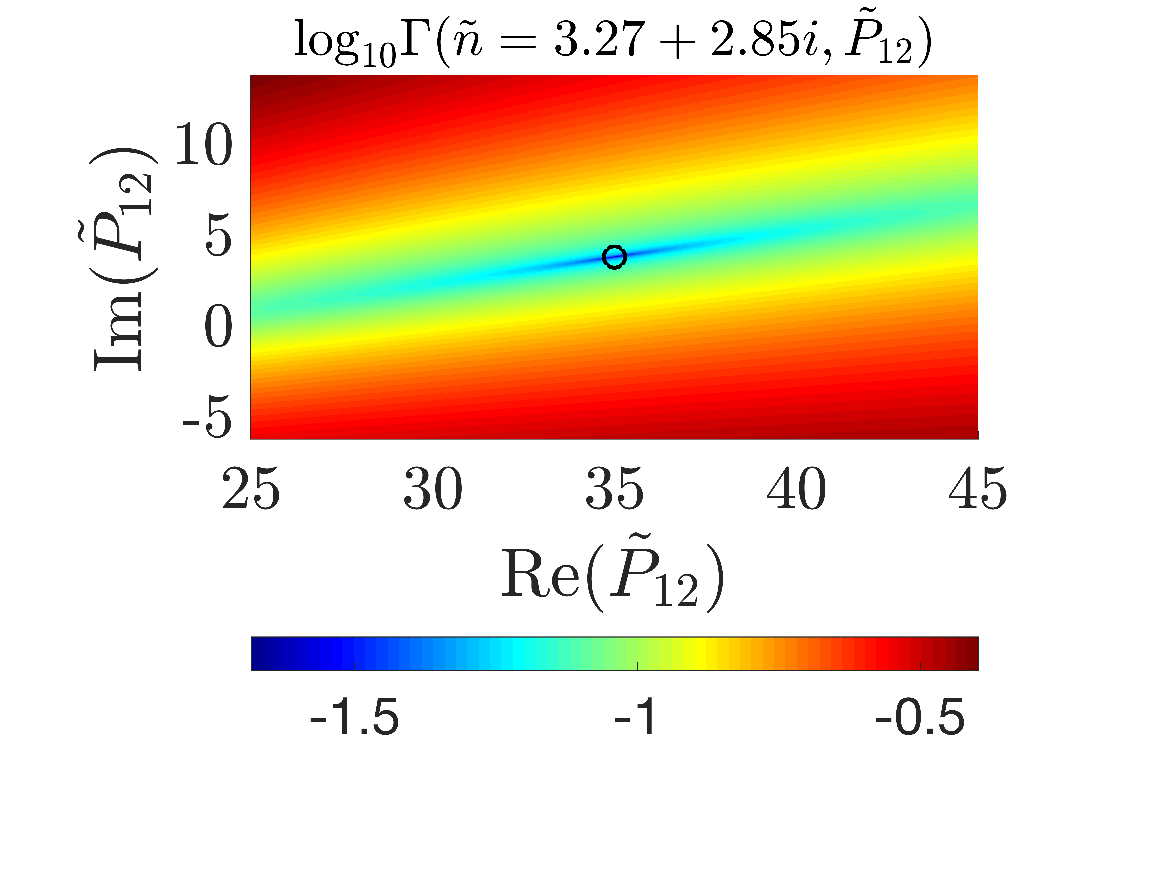}};
  \end{scope}
  \begin{scope}[overlay,shift={(90.8775948,-8.9742305)},scale=0.274874732,transform shape]
    \clip (119.5000,304.5000) rectangle (471.0000,353.7500);
    \node[anchor=north west,inner sep=0pt,outer sep=0pt] at (0,0)
      {\includegraphics[width=560bp]{Figs/residualOverP12ExpCr.eps}};
  \end{scope}
\end{tikzpicture}

\vspace*{-0.3cm}
\caption{Experimental real-strain residual landscapes for Cr near the selected solution. The residual $\Gamma$, defined in Eq.~\eqref{Alg4DIAG}, measures the normalized departure of the reconstructed strain spectrum from the real-strain condition. The left panel shows $\Gamma$ in the complex refractive-index plane, varying the real and imaginary parts of $\widetilde n$ with $\widetilde P_{12}$ fixed at the selected value. The right panel shows $\Gamma$ in the complex photoelastic-constant plane, varying the real and imaginary parts of $\widetilde P_{12}$ with $\widetilde n$ fixed at the selected value. The color scale shows $\log\Gamma$; the black circle marks the selected solution.}
\label{fig:GammaExpCr}
\end{figure}

Figure~\ref{fig:GammaExpCr} shows how the normalized real-strain residual
$\Gamma$, measuring the departure from the condition
$A(\omega)=A^{*}(-\omega)$, varies near the selected solution. The elongated
minima show that the real and imaginary parts of each material constant are
not constrained independently: a change in one can be partly compensated by
a change in the other while maintaining a nearly real strain spectrum.
Corresponding results for the synthetic data are shown in
Appendix~\ref{app:ToleranceLandscapes}.

Although $k_0$ and $\nu$ appear in Eq.~\eqref{echo1}, they enter only
through their product $\omega_0=k_0\nu=2\pi\nu/\lambda$. Equation~\eqref{echo1}
has an exact continuous scaling symmetry: for any real $\alpha>0$, the change
$\omega_0\rightarrow\alpha\omega_0$ combined with
$\widetilde n\rightarrow\widetilde n/\alpha$,
$\widetilde P_{12}\rightarrow
\widetilde P_{12}(\widetilde n^2-\alpha^2)/
[\alpha^2(\widetilde n^2-1)]$, and
$A(\omega)\rightarrow A(\omega)/\alpha$ leaves the calculated echo unchanged.
Thus the echo alone leaves one overall scaling degree of freedom; supplying
$\omega_0$ fixes this scale and is required to determine the individual
material constants and the absolute strain amplitude. The derivation is given
in Appendix~\ref{app:scaling}.

Once $\omega_0$ is supplied, the same measured echo generally corresponds
only to isolated alternative sets of material constants. A continuous family
of indistinguishable solutions can occur only when
\begin{equation}
\operatorname{Im}\left\{
(\widetilde P_{12}+\widetilde n^2-1)
[\widetilde n^2(\widetilde n^2-1)]^*
\right\}=0.
\label{eq:exceptionalcondition}
\end{equation}
For fixed refractive index $n$ and extinction coefficient $\kappa$, this
condition defines a straight line in the complex $\widetilde P_{12}$ plane;
when all four real material parameters vary, it forms an exceptional
hypersurface. Thus, except for this special case, the measured complex
transient provides independent constraints on the refractive index,
photoelastic constant, and strain. The derivation is given in
Appendix~\ref{app:scaling}.

\begin{figure}[H]
\centering
\vspace*{-0.25cm}
\hspace*{-0.65cm}
\begin{tikzpicture}[x=1bp,y=-1bp]
  \path[use as bounding box] (-23.19528,-37.00000) rectangle (220.00000,387.04124);
  \begin{scope}[overlay,shift={(-47.9527559,-11.5086614)},scale=0.365354331,transform shape]
    \clip (69.5000,19.7500) rectangle (450.7500,349.2500);
    \node[anchor=north west,inner sep=0pt,outer sep=0pt] at (0,0)
      {\includegraphics[width=560bp]{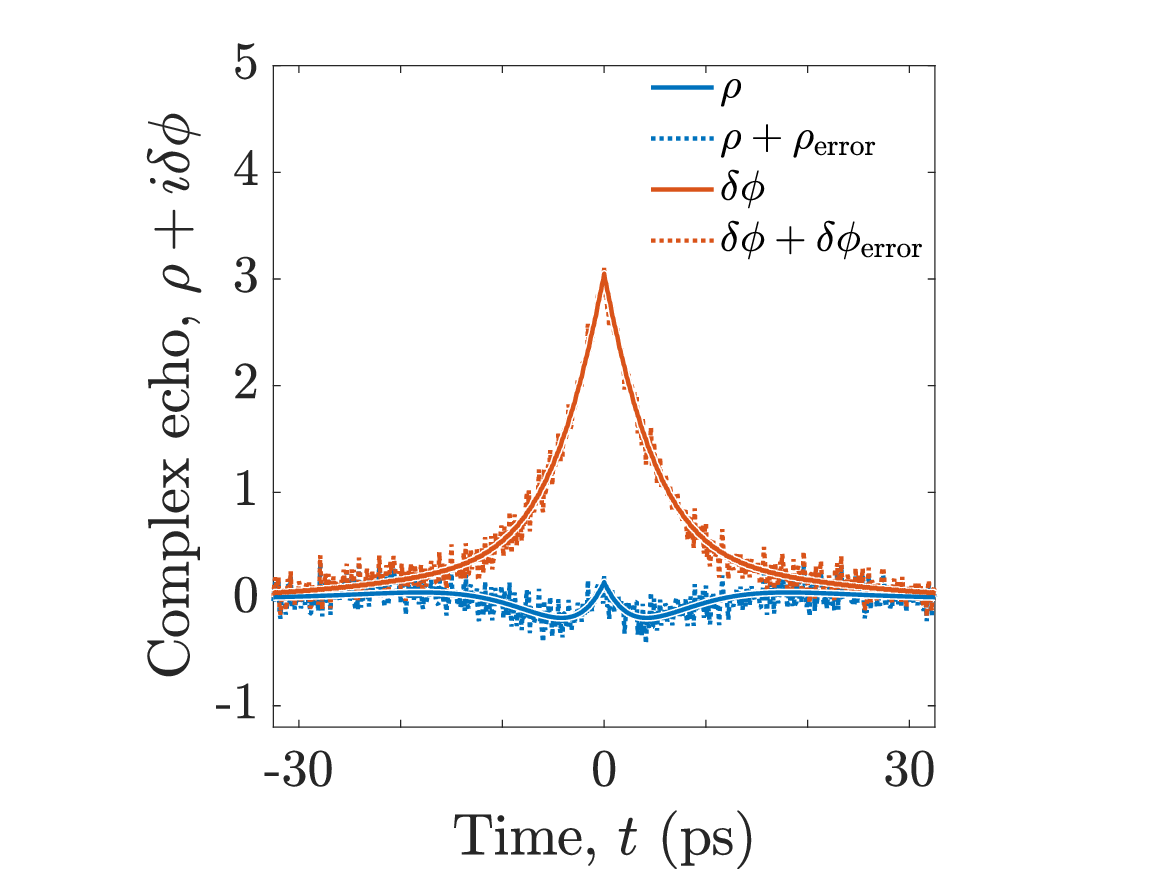}};
    \fill[white,overlay] (87.08871,152.00098) rectangle (96.28871,156.09100);
  \end{scope}
  \begin{scope}[overlay,shift={(-47.9527559,104.4913386)},scale=0.365354331,transform shape]
    \clip (69.5000,31.2500) rectangle (450.7500,418.2500);
    \node[anchor=north west,inner sep=0pt,outer sep=0pt] at (0,0)
      {\includegraphics[width=560bp]{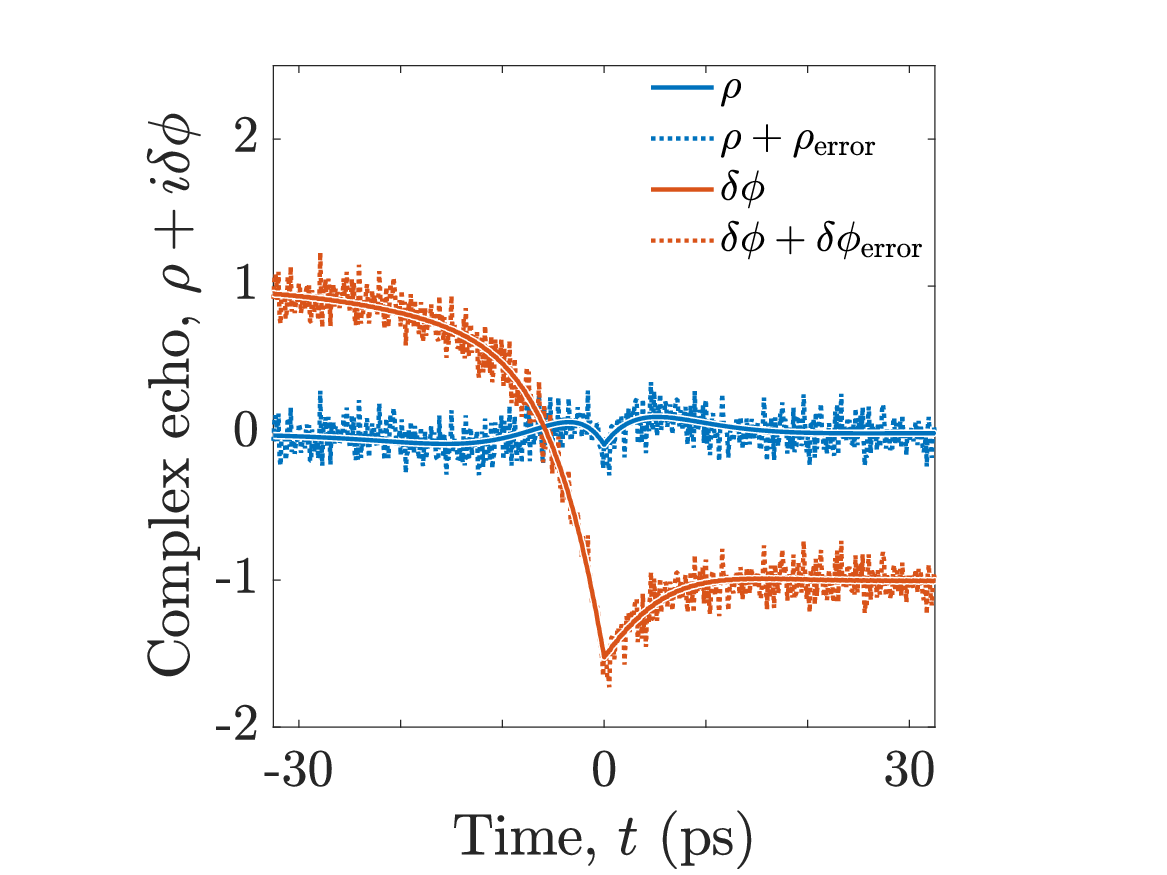}};
    \fill[white,overlay] (281.62101,406.67191) rectangle (285.71102,415.87188);
    \fill[white,overlay] (87.08871,152.00098) rectangle (96.28871,156.09100);
  \end{scope}
  \begin{scope}[overlay,shift={(56.6153846,-50.8615385)},scale=0.446153846,transform shape]
    \clip (148.0000,79.0000) rectangle (357.0000,374.2500);
    \node[anchor=north west,inner sep=0pt,outer sep=0pt] at (0,0)
      {\includegraphics[width=560bp]{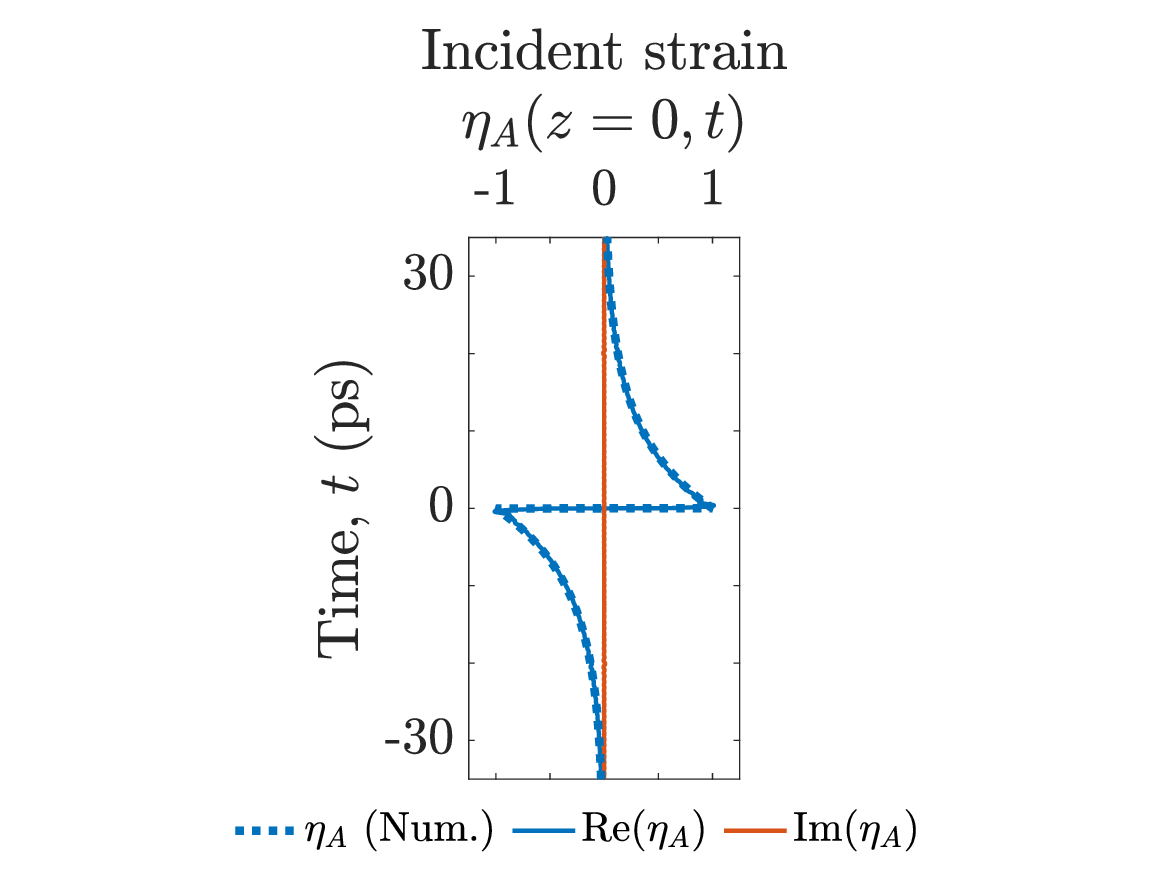}};
    \fill[white,overlay] (168.33901,248.28899) rectangle (177.53901,252.37901);
  \end{scope}
  \begin{scope}[overlay,shift={(56.6153846,65.1384615)},scale=0.446153846,transform shape]
    \clip (148.0000,113.7500) rectangle (357.0000,374.2500);
    \node[anchor=north west,inner sep=0pt,outer sep=0pt] at (0,0)
      {\includegraphics[width=560bp]{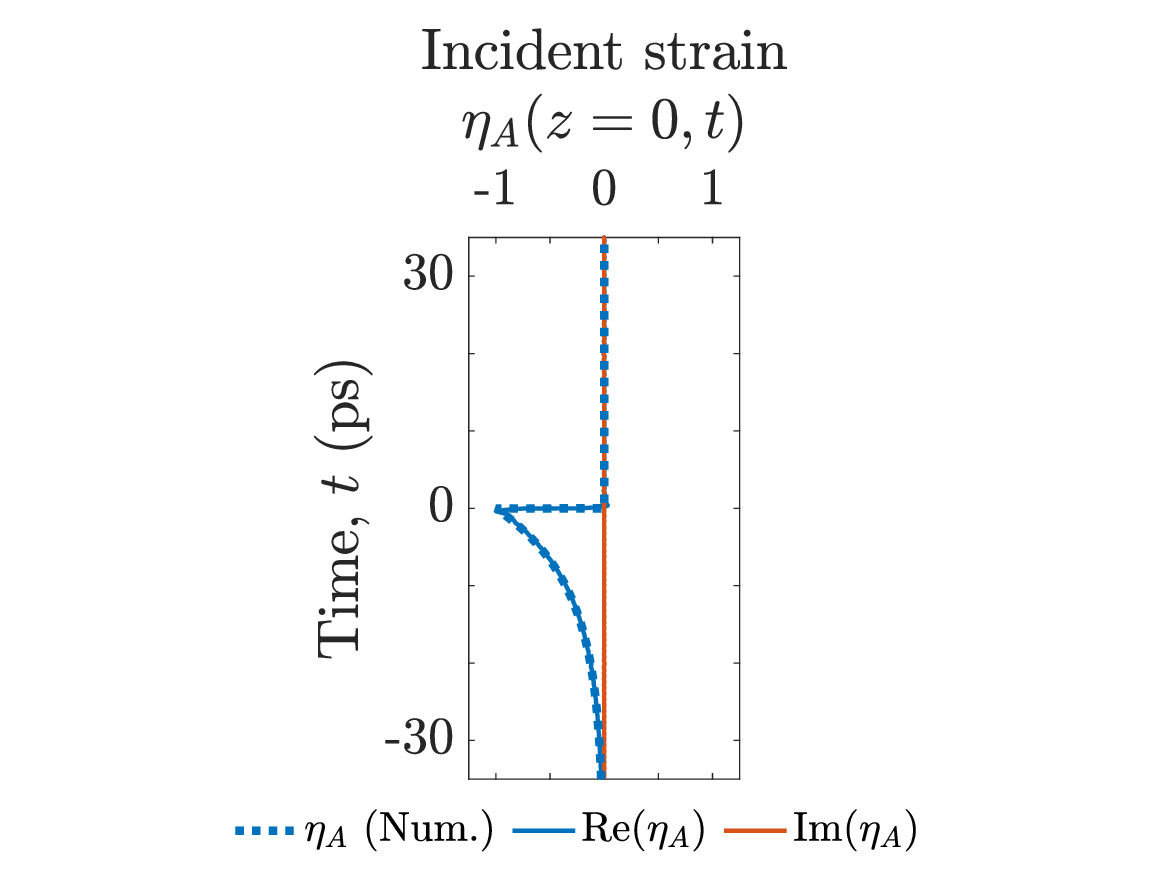}};
    \fill[white,overlay] (168.33901,248.28899) rectangle (177.53901,252.37901);
  \end{scope}
  \begin{scope}[overlay,shift={(80.0472441,-40.9322835)},scale=0.365354331,transform shape]
    \clip (198.0000,12.5000) rectangle (381.0000,34.5000);
    \node[anchor=north west,inner sep=0pt,outer sep=0pt] at (0,0)
      {\includegraphics[width=560bp]{Figs/etaABipoleAu.eps}};
  \end{scope}
  \begin{scope}[overlay,shift={(80.0472441,-43.4409449)},scale=0.365354331,transform shape]
    \clip (215.0000,44.0000) rectangle (365.0000,74.2500);
    \node[anchor=north west,inner sep=0pt,outer sep=0pt] at (0,0)
      {\includegraphics[width=560bp]{Figs/etaABipoleAu.eps}};
  \end{scope}
  \node[overlay,anchor=north east,inner sep=0pt,outer sep=0pt] at (245.2800,234.5000)
    {\includegraphics[trim={112bp 10bp 119bp 387.5bp},clip,width=120.201575bp]
      {Figs/etaABipoleAu.eps}};
  \begin{scope}[overlay,shift={(-23.3041475,229.1011521)},scale=0.456221198,transform shape]
    \clip (21.7500,78.5000) rectangle (510.7500,345.0000);
    \node[anchor=north west,inner sep=0pt,outer sep=0pt] at (0,0)
      {\includegraphics[width=560bp]{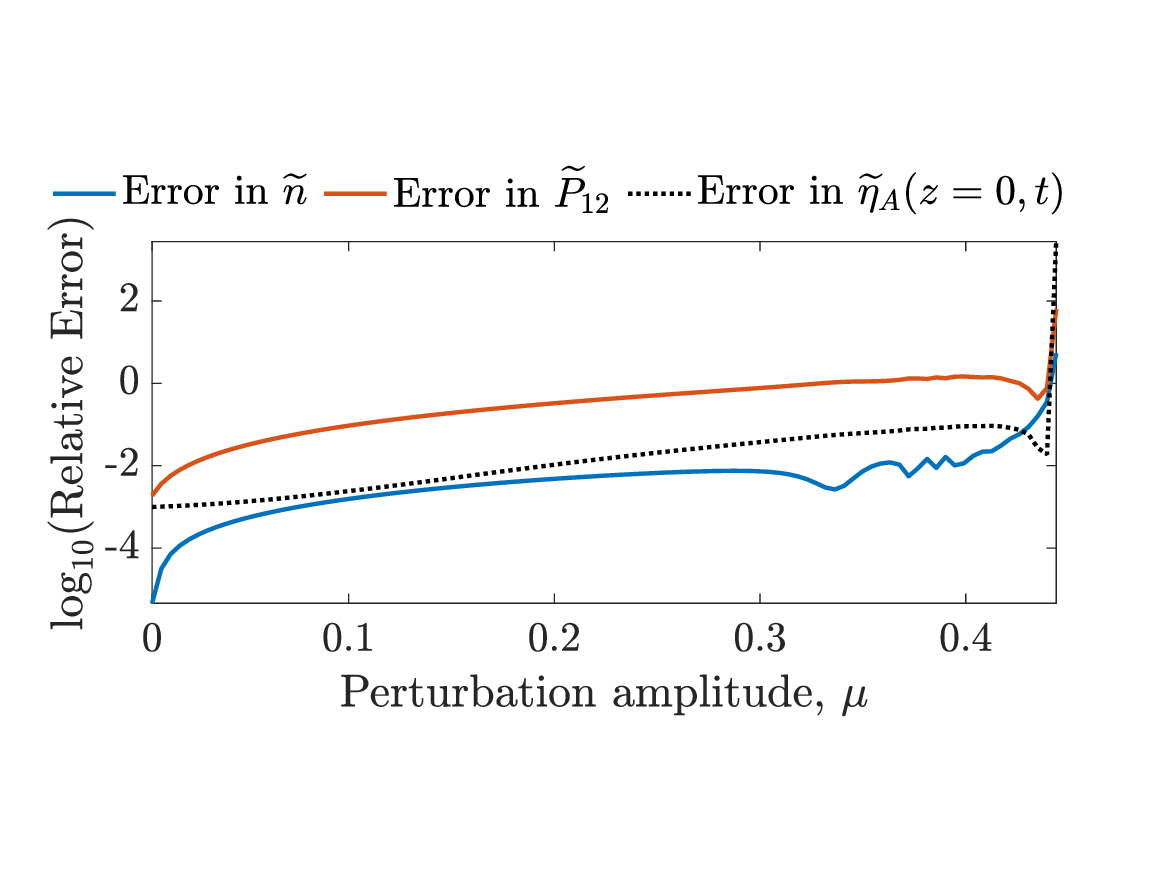}};
    \fill[white,overlay] (412.17100830,83.06199951) rectangle (423.87998657,86.00599365);
    \fill[white,overlay] (391.71900,336.25779) rectangle (395.09100,343.64608);
  \end{scope}
  \node[overlay,anchor=north west,inner sep=0pt,outer sep=0pt,
        font=\fontsize{9}{10}\selectfont] at (4.0000,4.0000) {$(a)$};
  \node[overlay,anchor=north west,inner sep=0pt,outer sep=0pt,
        font=\fontsize{9}{10}\selectfont] at (4.0000,120.0000) {$(b)$};
  \node[overlay,anchor=north west,inner sep=0pt,outer sep=0pt,
        font=\fontsize{9}{10}\selectfont] at (161.0000,4.0000) {$(c)$};
  \node[overlay,anchor=north west,inner sep=0pt,outer sep=0pt,
        font=\fontsize{9}{10}\selectfont] at (161.0000,120.0000) {$(d)$};
  \node[overlay,anchor=north west,inner sep=0pt,outer sep=0pt,
        font=\fontsize{9}{10}\selectfont] at (16.0000,287.0000) {$(e)$};
\end{tikzpicture}
\vspace*{-0.25cm}
\caption{Calculated time-domain data for the synthetic pulses.
Panels (a) and (b) show $\rho(t)$ and $\delta\phi(t)$ for the bipolar and
unipolar strain pulses, respectively, for $\mu=0$ and $0.1$.
Panels (c) and (d) show the input incident strain waveforms and the
real and imaginary parts reconstructed for $\mu=0$.
Panel (e) shows the logarithms of the relative errors $|x_{\mathrm{rec}}-x_0|/|x_0|$
for $x=\widetilde n$ and $\widetilde P_{12}$, together with the
normalized strain error
$E_\eta=[\sum_t|\eta_{A,\mathrm{rec}}(t)-\eta_{A,0}(t)|^2/
\sum_t|\eta_{A,0}(t)|^2]^{1/2}$,
versus $\mu$ for the bipolar pulse, where the subscript $0$ denotes the
values used to generate the synthetic data.}
\label{fig:refStrainAu}
\end{figure}

We next test the reconstruction using known material constants and strain pulses, and examine its sensitivity to added noise. For a probe wavelength $\lambda=400$~nm, we use literature
values for gold: $\widetilde n=1.47+1.95i$, $\widetilde P_{12}=1.98+10.74i$,
$\nu=3300$~m~s$^{-1}$, and $\zeta_p=32$~nm
\cite{johnson1972optical,pezeril2014femtosecond}. We prescribe bipolar and
unipolar strain pulses, shown in Fig.~\ref{fig:refStrainAu}(c) and (d),
respectively. The bipolar incident waveform at the surface is
$\eta_A(t)=\operatorname{sgn}(t)\exp[-\nu|t|/\zeta_p]$.
The unipolar pulse is identical to the bipolar pulse for $t<0$ and vanishes for $t>0$.

To test robustness to noise, we perturb the calculated $\delta r(t)/ r$ by adding $\mu(1+i)\,\mathrm{randn}(t)$, where $\mathrm{randn}(t)$ is unit-standard-deviation Gaussian white noise and $\mu$ its amplitude. For both pulse shapes, at $\mu=0$ the errors
in $\widetilde n$ and $\widetilde P_{12}$ are below $0.19\%$, and the
reconstructed strain waveforms closely match the input strain waveforms, with
negligible imaginary parts, as shown in Fig.~\ref{fig:refStrainAu}(c),(d). At
$\mu=0.1$, the corresponding errors remain below $0.2\%$ and $10\%$,
respectively. Figure~\ref{fig:refStrainAu}(e) shows that the normalized
strain-reconstruction error increases from about $0.1\%$ at $\mu=0$ to
about $8\%$ at $\mu=0.4$. For $0\lesssim\mu\lesssim0.4$ the
$\widetilde P_{12}$ error is typically one to two orders of magnitude larger
than the $\widetilde n$ error, and both diverge for $\mu\gtrsim0.44$. The close agreement between the reconstructed and input waveforms for both
pulse shapes shows that the reconstruction does not require an assumed
temporal symmetry. Further results, including the corresponding unipolar
error curves, are given in the Supplemental Material.

Joint recovery of the strain profile and material response allows
acoustic attenuation, thermal diffusion, and photoexcited carrier diffusion \cite{Thomsen1986Surface,wright1994,wright2001ultrafast} to be investigated
without independently supplied optical or photoelastic constants, making the
method particularly useful for metals and semiconductors.
Repeating the experiment at different probe wavelengths could determine the
spectral dependence of the complex dielectric and photoelastic constants,
including the contribution of electronic transitions to the complex
photoelastic response.

At heart, the problem resembles a snake biting its own tail: recovering the strain field requires knowledge of the material properties through which it is observed, whereas determining those properties requires knowledge of the field. In contrast to ptychographic diffractive imaging, where redundancy is introduced through a series of diffraction measurements from spatially overlapping illuminated regions, here broadband temporal information within a single complex echo resolves this mutual dependence. No spatial scanning or externally imposed diversity in incidence angle, probe wavelength, polarization, or illumination is required for the local reconstruction, making the measurement self-referencing. 

Scanning the focused optical spot across the sample would allow point-by-point determination of the complex refractive index $\widetilde n$, complex photoelastic constant $\widetilde P_{12}$, and strain waveform, with a lateral resolution potentially approaching $\sim10~\mu\mathrm{m}$. This simultaneous local recovery combines optical, photoelastic, and mechanical information that is otherwise obtained separately. The present approach therefore opens the way to two-dimensional imaging of optical, photoelastic, and mechanical properties in heterogeneous materials and microstructured devices at normal optical incidence, allowing spatial variations in dielectric response, photoelastic coupling, and strain dynamics to be correlated directly.

The same principle could apply whenever a time-dependent field produces both surface motion and a change in optical properties. A temperature field, for example, produces surface motion through thermal expansion and changes the dielectric response through the thermo-optic effect, so that the temperature field, complex refractive index, and thermo-optic constants could in principle be determined together. Photoexcited carrier distributions can similarly generate strain through deformation-potential stress while changing the refractive index, allowing the carrier-density field, complex refractive index, and carrier-density dependence of the refractive index to be determined together. In these and other systems, on any time scale for which both effects can be measured and the coupling between the field and surface motion is known, that motion can provide the internal reference needed to determine the field and the material-dependent optical properties through which it is observed.$\,$

$\,$

\begin{acknowledgments}
\noindent R.W. acknowledges funding from the EU H2020 projects MetaVEH (grant agreement No. 952039) and DYNAMO (grant agreement No. 101046489). O.B.W., H.O., O.M., and M.T. acknowledge Grants-in-Aid for Scientific
Research from the Ministry of Education, Culture, Sports, Science, and
Technology (MEXT), Japan.
\end{acknowledgments}

$ \, $

$ \, $


\noindent \textbf{Code availability:} The experimental Cr data \cite{saito2003picosecond,saito2004erratum} and MATLAB code implementing Algorithm~2 of the Supplemental Material are provided as ancillary files accompanying this paper. These files recompute the experimental reconstructions in Figs.~\ref{fig:BSwithReflectanceCr}, \ref{fig:RaAaEtaACr}, and~\ref{fig:CrTimeRecovery}, as well as the singular strain-spectrum example in Fig.~S1 of the Supplemental Material.

\vspace{-0.5cm}
%

\appendix

\section{Structure-exploiting Newton iteration}
\label{APPEN}

We now describe the numerical procedure used to determine the material
constants from Eq.~\eqref{materialconstraint}. Write
$\widetilde P_{12}=P\exp(i\theta)$ and
$\widetilde n=N\exp(i\psi)$, where $P$, $\theta$, $N$, and $\psi$ are real.
For the $m$ retained frequencies $\omega_q$, define the diagonal matrix
$T(P,\theta,N,\psi)$ by
$T_{qq}=J(\omega_q)G(\omega_q)+G^*(\omega_q)$, where $J$ and $G$ are defined
in the main text, and let $\mathbf A$ contain the corresponding values
$A(\omega_q)$. Equation~\eqref{materialconstraint} then requires the same
four material parameters to make the relevant diagonal entries of $T$ vanish
at all retained frequencies. Evaluating Eq.~\eqref{materialconstraint} multiplied by $A(\omega_q)$ at the retained frequencies yields the matrix equation
\begin{equation}
    T(P,\theta,N,\psi)\mathbf{A} = \mathbf{0} \, . \label{NLEVP}
\end{equation}
Eq.~\eqref{NLEVP} is a multiparameter nonlinear eigenvalue problem
\cite{guttel2017nonlinear,voss2014nonlinear,vial2024quasinormal,marti2025fabry}. The material constants are the
nonlinear eigenvalues and the broadband strain spectrum the eigenvector.
The derivation of Eqs.~\eqref{materialconstraint} and \eqref{NLEVP} is given in the
Supplemental Material.

Several variants of generalized Rayleigh-quotient iteration are available
\cite{schreiber2008nonlinear,guttel2017nonlinear}. For the present diagonal
problem, however, the finite-iterate eigenvectors of the operator are
generically coordinate-basis vectors when the diagonal entries are distinct,
or belong to degenerate coordinate subspaces when diagonal entries coincide,
rather than representing the broadband data-constrained strain spectrum.
We therefore use a structure-exploiting procedure in which the strain spectrum
is reconstructed directly from the measured echo after each material-parameter
update, and convergence is assessed through its physical conjugate symmetry.
The standard generalized Rayleigh-quotient iteration and its relation to the
present procedure are given in the Supplemental Material. For comparison, the standard iteration in $P$, with $\theta$, $N$, and $\psi$ held fixed, assesses convergence through
\begin{equation}
\left\|
T(P^{(\tau)},\theta,N,\psi)
\mathbf{A}_{R}^{(\tau)}
\right\|
<\Gamma_{\mathrm{std}}.
\label{Alg4DENSE}
\end{equation}

\subsection{Common-zero structure of the diagonal problem}
\label{CommonZeroStructure}

For an exact solution with $ A(\omega_q)\neq0$,
Eq.~\eqref{NLEVP} requires
$T_{qq}(P,\theta,N,\psi)=0$ for all $q=1,\ldots,m$.
The eigensolution is therefore a common zero of all diagonal entries,
a stronger condition than $\det T=0$, which requires only one diagonal entry to vanish.
At this common zero, $T=0$, so Eq.~\eqref{NLEVP} no longer determines
$\mathbf A$. Accordingly, we locate the common zero by Newton iteration
within a modified generalized Rayleigh-quotient procedure, reconstructing
$\mathbf A$ directly from Eq.~\eqref{echo1} after each update of
$(P,\theta,N,\psi)$.

Convergence is assessed by a residual $\Gamma$
(see Eqn.~\eqref{Alg4DIAG} and Supplemental Material), which measures the
departure of the reconstructed spectrum from
$ A(\omega)= A^{*}(-\omega)$. The conditions
$T_{qq}=0$ constrain $\widetilde n$ and $\widetilde P_{12}$ without an
assumed strain profile, while Eq.~\eqref{echo1} reconstructs the associated
spectrum directly from the measured $ R(\omega)$.

At the exact common zero,
\begin{equation}
T_{qq}(P,\theta,N,\psi)=0,
\qquad q=1,\ldots,m,
\label{ExactCommonZero}
\end{equation}
so $T(P,\theta,N,\psi)$ is the zero matrix and its nullspace is the full
$m$-dimensional space. Equation~\eqref{NLEVP} alone therefore cannot select a
unique $\mathbf A$. The measured echo removes this ambiguity: once
$\widetilde n$ and $\widetilde P_{12}$ are known, Eq.~\eqref{echo1} uniquely
reconstructs every retained $ A(\omega_q)$ from the measured
$ R(\omega_q)$.

At a finite iterate the common-zero condition is not yet satisfied. The
procedure described below therefore reconstructs a data-constrained trial
eigenvector from the current material-parameter estimate after every update.
At convergence this reconstruction selects, from the full nullspace of the
zero matrix, the strain spectrum associated with the measured echo. The
iteration thus targets the common-zero structure in
Eq.~\eqref{ExactCommonZero}, rather than determinant singularity at an
individual frequency.

\subsection{Newton's method with a modified generalized Rayleigh quotient}
\label{ModRayleighQuotient}

At each iterate, with the independently supplied frequency scale
$k_{0}\nu$ fixed, we first reconstruct the data-constrained trial vector
directly from the measured spectrum. For given $P^{(\tau)}$, $\theta^{(\tau)}$,
$N^{(\tau)}$, and $\psi^{(\tau)}$, Eq.~\eqref{echo1} gives
\begin{small}
\begin{equation}
\begin{split}  
 {A}(\omega_{q}) = \quad \quad  \quad \quad \quad \quad  \quad \quad  \quad \quad  \quad \quad  \quad \quad  \quad \quad  \quad \quad  \quad \quad  \quad \\
 = \frac{{R}(\omega_{q})}{4i k_{0} \nu \left\lbrace \frac{\omega_{q}^{2} P^{(\tau)} \exp(i \theta^{(\tau)}) }{(N^{(\tau)2} \exp(2i \psi^{(\tau)})  - 1)(\omega_{q}^{2} - 4 N^{(\tau)2} \exp(2i \psi^{(\tau)})  k_{0}^{2} \nu^{2})} + 1 \right\rbrace} . 
\end{split} 
 \label{Alg1DIAG}
\end{equation}
\end{small}
Equation~\eqref{Alg1DIAG} determines every component of the
data-constrained trial vector $\mathbf{A}^{(\tau)}$ from the current
material-parameter estimate and the measured data. When this vector enters
the generalized-Rayleigh construction, we take
\begin{equation}
\mathbf{A}_{R}^{(\tau)}
=
\mathbf{A}_{L}^{(\tau)}
=
\mathbf{A}^{(\tau)} .
\label{Alg2DIAG}
\end{equation}
For evenly spaced, origin-symmetric samples $\{\omega_q\}_{q=1}^{m}$, we
replace Eq.~\eqref{Alg4DENSE} by the normalized real-strain residual
\begin{equation}
\Gamma=
\frac{
\left\|
W\left(\mathbf A^{(\tau)}
-\mathrm{flip}(\mathbf A^{(\tau)})^{*}\right)
\right\|
}{
\left\|W\mathbf A^{(\tau)}\right\|
},
\label{Alg4DIAG}
\end{equation}
where $\mathrm{flip}$ reverses the order of the frequency samples and
$W=\mathrm{diag}\left(
\| R(\omega_1)\|,
\| R(\omega_2)\|,\ldots,
\| R(\omega_m)\|\right)$.
The numerator vanishes when
$ A(\omega)= A^{*}(-\omega)$.
The weighting de-emphasises frequency bins with small
$\| R(\omega_q)\|$, for which the reconstructed phase and
amplitude are less reliable \cite{tarantola2005inverse}.
Numerical roots satisfying $\Gamma<\Gamma_{\mathrm{min}}$ are retained
as admissible solutions, with $\Gamma_{\mathrm{min}}$ denoting the
acceptance threshold.

To update all four parameters simultaneously, we partition
the diagonal system into four blocks:
\begin{widetext}
\begin{equation}
T(P,\theta,N,\psi)\mathbf{A}
=
\begin{pmatrix}
T_{1}(P,\theta,N,\psi) & & & \\
& T_{2}(P,\theta,N,\psi) & & \\
& & T_{3}(P,\theta,N,\psi) & \\
& & & T_{4}(P,\theta,N,\psi)
\end{pmatrix}
\begin{bmatrix}
\mathbf{A}_1\\
\mathbf{A}_2\\
\mathbf{A}_3\\
\mathbf{A}_4
\end{bmatrix}
=
\mathbf{0}.
\end{equation}
\end{widetext}
The four blocks provide four scalar parameter conditions. Define
\begin{equation}
\mathcal{N}
\begin{bmatrix}
P\\
\theta\\
N\\
\psi\\
\mathbf{A}
\end{bmatrix}
=
\begin{bmatrix}
\mathbf{A}_1^{\dagger}
T_1(P,\theta,N,\psi)\mathbf{A}_1\\
\mathbf{A}_2^{\dagger}
T_2(P,\theta,N,\psi)\mathbf{A}_2\\
\mathbf{A}_3^{\dagger}
T_3(P,\theta,N,\psi)\mathbf{A}_3\\
\mathbf{A}_4^{\dagger}
T_4(P,\theta,N,\psi)\mathbf{A}_4\\[0.3em]
T_1\mathbf{A}_1\\
T_2\mathbf{A}_2\\
T_3\mathbf{A}_3\\
T_4\mathbf{A}_4
\end{bmatrix}
=
\mathbf{0}.
\label{NewtonVector}
\end{equation}
The four scalar conditions in Eq.~\eqref{NewtonVector} imply
$\mathbf{A}^{\dagger}T(P,\theta,N,\psi)\mathbf{A}=0$, since
\begin{equation}
\begin{split}
\mathbf{A}^{\dagger}T(P,\theta,N,\psi)\mathbf{A}
={}&
\mathbf{A}_1^{\dagger}T_1\mathbf{A}_1
+
\mathbf{A}_2^{\dagger}T_2\mathbf{A}_2
\\
&+
\mathbf{A}_3^{\dagger}T_3\mathbf{A}_3
+
\mathbf{A}_4^{\dagger}T_4\mathbf{A}_4 .
\end{split}
\label{actualRQiterrate}
\end{equation}
The converse does not follow from the summed condition in
Eq.~\eqref{actualRQiterrate} alone, because cancellation between the four
terms is possible. When $m$ is divisible by four, each block contains
$m/4$ frequency samples. At the desired common-zero solution,
Eq.~\eqref{ExactCommonZero} makes every diagonal entry of each block vanish,
so that $T_r\mathbf{A}_r=\mathbf{0}$ and hence
$\mathbf{A}_r^{\dagger}T_r\mathbf{A}_r=0$ for $r=1,\ldots,4$. Thus the four block conditions vanish separately at the
desired common-zero solution.

A first-order expansion of Eq.~\eqref{NewtonVector} defines an augmented
Jacobian $J_N$ and gives a Newton step that updates the four material
parameters $P$, $\theta$, $N$, and $\psi$ together. After each
material-parameter update, however, the data-constrained trial vector is
reconstructed directly from Eq.~\eqref{Alg1DIAG}, rather than retaining the
vector components generated by the Newton step.

At the exact common-zero solution, $T(P,\theta,N,\psi)=0$ and the full
augmented Jacobian is singular, so $J_N$ may become poorly conditioned as the
solution is approached. The calculation therefore terminates at the finite
conjugate-symmetry tolerance in Eq.~\eqref{Alg4DIAG}. When inversion of the
full augmented Jacobian becomes unreliable, the four material parameters may
instead be updated sequentially. For computational efficiency, initial guesses are screened using the conjugate-symmetry residual before Newton refinement. The explicit Newton update, constituent blocks of $J_N$, the screening criterion, and both iteration algorithms are given in the Supplemental Material.

\section{Real-strain residual landscapes}
\label{app:ToleranceLandscapes}

\begin{figure}[H]
\vspace*{-0.25cm}
\hspace*{-0.6cm}
\begin{tikzpicture}[x=1bp,y=-1bp,scale=1.1, transform shape]
  \path[use as bounding box] (-21.57143,-10.00215) rectangle (224.77953,89.05655);
  \begin{scope}[overlay,shift={(-29.4725275,-9.2307692)},scale=0.263736264,transform shape]
    \clip (32.7500,1.2500) rectangle (490.7500,299.2500);
    \node[anchor=north west,inner sep=0pt,outer sep=0pt] at (0,0)
      {\includegraphics[width=560bp]{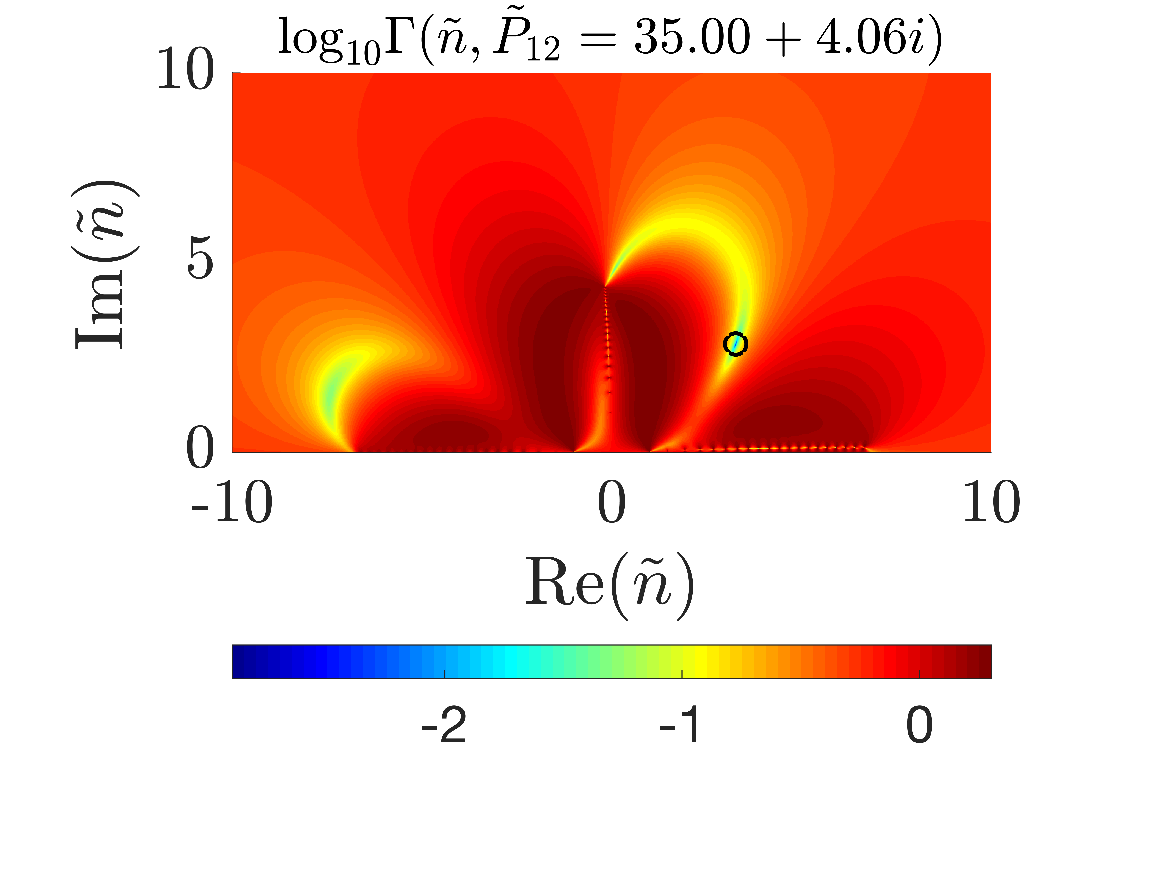}};
  \end{scope}
  \begin{scope}[overlay,shift={(-29.4725275,-6.5604396)},scale=0.263736264,transform shape]
    \clip (110.5000,308.2500) rectangle (477.0000,357.7500);
    \node[anchor=north west,inner sep=0pt,outer sep=0pt] at (0,0)
      {\includegraphics[width=560bp]{Figs/residualOverNExactCr.eps}};
  \end{scope}
  \begin{scope}[overlay,shift={(90.8775948,-9.8954903)},scale=0.274874732,transform shape]
    \clip (36.0000,2.2500) rectangle (484.2500,295.5000);
    \node[anchor=north west,inner sep=0pt,outer sep=0pt] at (0,0)
      {\includegraphics[width=560bp]{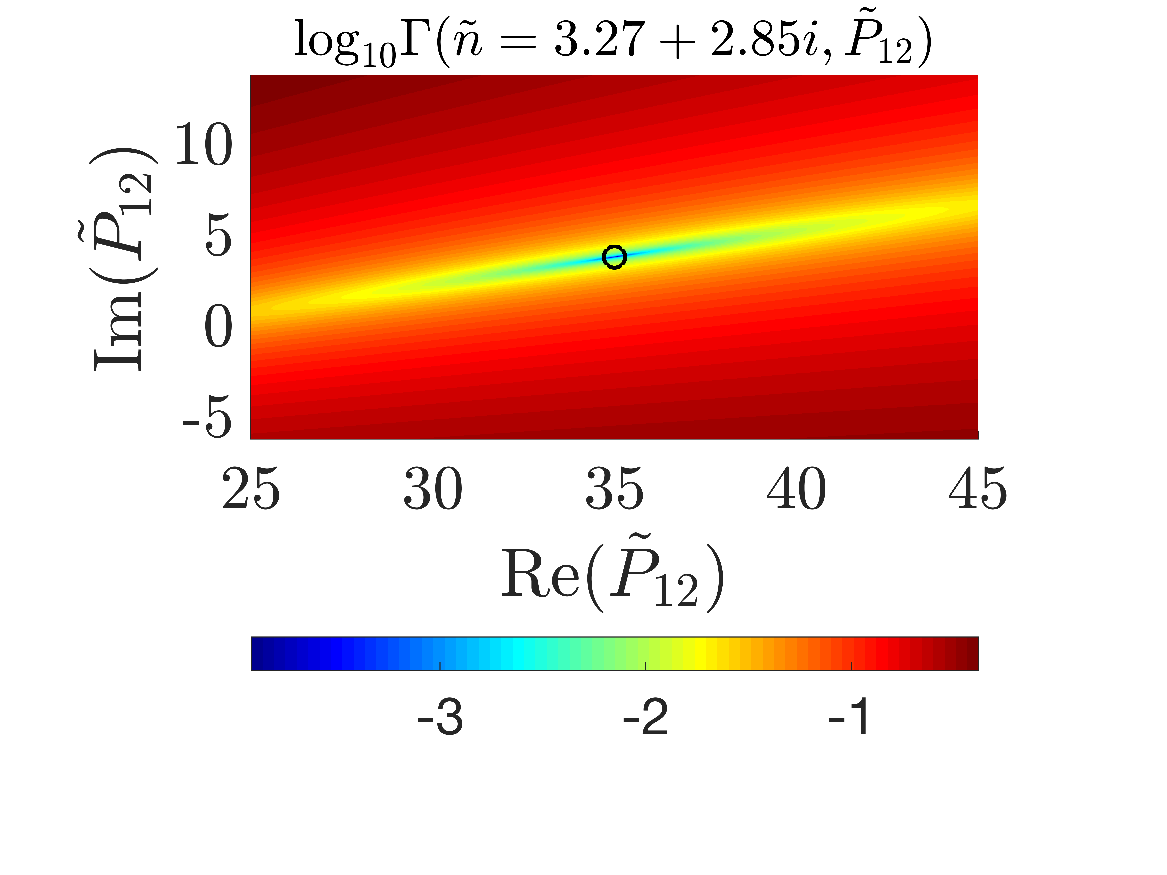}};
  \end{scope}
  \begin{scope}[overlay,shift={(90.8775948,-8.9742305)},scale=0.274874732,transform shape]
    \clip (119.5000,304.5000) rectangle (471.0000,353.7500);
    \node[anchor=north west,inner sep=0pt,outer sep=0pt] at (0,0)
      {\includegraphics[width=560bp]{Figs/residualOverP12ExactCr.eps}};
  \end{scope}
\end{tikzpicture}
\vspace*{-0.6cm}
\caption{Calculated real-strain residual landscapes for Cr near the
selected solution. The residual $\Gamma$ measures the normalized departure
from the real-strain condition $A(\omega)=A^{*}(-\omega)$. The left panel
shows $\Gamma$ in the complex refractive-index plane, varying the real and
imaginary parts of $\widetilde n$ with $\widetilde P_{12}$ fixed at the
selected value. The right panel shows $\Gamma$ in the complex
photoelastic-constant plane, varying the real and imaginary parts of
$\widetilde P_{12}$ with $\widetilde n$ fixed at the selected value. The
color scale shows $\log\Gamma$; the black circle marks the selected solution.}
\label{app:fig:logGammaColourPlots}
\end{figure}

The experimental real-strain residual landscapes in the complex
refractive-index and complex photoelastic-constant planes are shown in
Fig.~\ref{fig:GammaExpCr} in the main text. Figure~\ref{app:fig:logGammaColourPlots}
shows the corresponding landscapes calculated from the reconstructed echo.
The calculated landscapes show similar elongated minima and place the
selected solution near their minima, in qualitative agreement with experiment.
The experimental landscapes have larger values of $\Gamma$, reflecting their
greater departure from the real-strain condition.

\section{Dimensionless structure and ambiguities}
\label{app:scaling}

There are two distinct ways in which different material parameters and strain
profiles can give the same complex echo. The first is the exact continuous
scaling symmetry given in the main text. Supplying $\omega_0$ independently
removes this ambiguity.

To expose the remaining structure, we write Eq.~\eqref{echo1} in
dimensionless form. Defining
$\Omega=\omega/\omega_0$,
$\mathcal{R}(\Omega)=R(\omega_0\Omega)$, and
$\mathcal{A}(\Omega)=\omega_0A(\omega_0\Omega)$ gives
\begin{equation}
\mathcal{R}(\Omega)
=
4i
\left\{
1+
\frac{\Omega^{2}\widetilde P_{12}}
{(\widetilde n^{2}-1)
(\Omega^{2}-4\widetilde n^{2})}
\right\}
\mathcal{A}(\Omega).
\label{dimensionlessEcho}
\end{equation}
Here $\omega_0$ no longer appears explicitly, and the reality condition becomes
$\mathcal A(-\Omega)=\mathcal A^*(\Omega)$. The product
$\widetilde n\omega_0$ is invariant under the scaling transformation and
also gives the normal-incidence Brillouin angular frequency through
$\omega_B=2\,\mathrm{Re}(\widetilde n\omega_0)$.

Even after $\omega_0$ is supplied, a further ambiguity in the material
parameters can remain. In general this is at most a discrete alternative, but
for special values of the material constants it becomes a continuous family.
To identify these cases, define
\begin{equation}
d=
\frac{\widetilde P_{12}+\widetilde n^2-1}
{\widetilde n^2(\widetilde n^2-1)}.
\label{dDefinition}
\end{equation}
Equation~\eqref{dimensionlessEcho} can then be written as
\begin{equation}
\mathcal R(\Omega)=4iG(\Omega)\mathcal A(\Omega),
\qquad
G(\Omega)=
\frac{1-d\Omega^2/4}
{1-\Omega^2/(4\widetilde n^2)}.
\label{dimensionlessG}
\end{equation}
The material dependence of $G(\Omega)$ is therefore expressed through
$\widetilde n$ and the single combination $d$.

The reality of the strain allows $\mathcal A$ to be eliminated by comparing
positive and negative frequencies. For $\Omega>0$, define
\begin{equation}
H(\Omega)\equiv
-\frac{\mathcal R(\Omega)}
{\mathcal R^*(-\Omega)}.
\label{phaseRelationDefinition}
\end{equation}
Using $\mathcal A(-\Omega)=\mathcal A^*(\Omega)$ gives
\begin{equation}
H(\Omega)=\frac{G(\Omega)}{G^*(\Omega)}
=
\frac{(1-d\Omega^2/4)
[1-\Omega^2/(4(\widetilde n^*)^2)]}
{(1-d^*\Omega^2/4)
[1-\Omega^2/(4\widetilde n^2)]}.
\label{phaseRelation}
\end{equation}
The unknown strain spectrum has therefore disappeared, whereas the
frequency dependence retains the material information.

For an absorbing medium, $\widetilde n^2$ has a nonzero imaginary part. When
$\operatorname{Im}d\neq0$ and there is no numerator--denominator
cancellation, $H(\Omega)$ determines the two distinct complex coefficients
$d$ and $1/(\widetilde n^*)^2$ appearing in its numerator. These two
coefficients can be assigned to the photoelastic numerator and the optical
pole in two ways. One assignment reproduces the original material
parameters. The only other possibility exchanges their roles. Denoting this
alternative solution by $\widetilde n'$ and $\widetilde P_{12}'$, with $d'$
defined as in Eq.~\eqref{dDefinition}, the exchange requires
$d'=1/(\widetilde n^*)^2$ and
$1/(\widetilde n')^2=d^*$, corresponding to
$G'(\Omega)=1/G^*(\Omega)$. Thus the general case has at most one additional
discrete reciprocal-conjugate solution rather than a continuous family. For
the synthetic material parameters considered below, this alternative is
nonpassive and is therefore excluded on physical grounds.

A continuous ambiguity arises when $d$ is real. In this case the factor
$1-d\Omega^2/4$ is real and cancels completely from
Eq.~\eqref{phaseRelation}, leaving
\begin{equation}
H(\Omega)=
\frac{1-\Omega^2/(4(\widetilde n^*)^2)}
{1-\Omega^2/(4\widetilde n^2)}.
\label{exceptionalH}
\end{equation}
The complex echo still fixes the complex optical pole, but the real numerator
factor can no longer be determined. For every real $\beta$,
\begin{equation}
\widetilde n_\beta=\widetilde n,\qquad
\widetilde P_{12,\beta}
=(\beta\widetilde n^2-1)(\widetilde n^2-1),
\label{exceptionalPfamily}
\end{equation}
and
\begin{equation}
G_\beta(\Omega)=
\frac{1-\beta\Omega^2/4}
{1-\Omega^2/(4\widetilde n^2)},
\qquad
\mathcal A_\beta(\Omega)=
\frac{1-d\Omega^2/4}
{1-\beta\Omega^2/4}\,
\mathcal A(\Omega)
\label{exceptionalFamily}
\end{equation}
produce exactly the same complex echo. The original solution corresponds to
$\beta=d$.

Equation~\eqref{exceptionalPfamily} shows directly that, for fixed
$\widetilde n$, these exceptional values form a straight line in the complex
$\widetilde P_{12}$ plane. Eliminating the auxiliary parameter $\beta$ gives
the compact condition quoted in the main text,
\begin{equation}
\operatorname{Im}
\left\{
(\widetilde P_{12}+\widetilde n^2-1)
\left[\widetilde n^2(\widetilde n^2-1)\right]^*
\right\}=0.
\label{exceptionalCompact}
\end{equation}
Writing $\widetilde n=n+i\kappa$ and
$\widetilde P_{12}=P_R+iP_I$, Eq.~\eqref{exceptionalCompact} becomes
\begin{equation}
\begin{split}
&(P_I+2n\kappa)
\left[(n^2-\kappa^2)(n^2-\kappa^2-1)-4n^2\kappa^2\right] \\
&\qquad -2n\kappa[2(n^2-\kappa^2)-1]
(P_R+n^2-\kappa^2-1)=0.
\end{split}
\label{eq:exceptionalconditionExpanded}
\end{equation}
For fixed refractive index $n$ and extinction coefficient $\kappa$,
Eq.~\eqref{eq:exceptionalconditionExpanded} defines a straight line in the complex
$\widetilde P_{12}$ plane. When all four real material parameters vary, this
single exact relation defines an exceptional hypersurface in the full
parameter space. Thus, away from this special set, no continuous ambiguity of
this type occurs.

This exceptional condition is independent of the scaling ambiguity described
in the main text. Under that transformation,
\begin{equation}
\Omega\rightarrow\frac{\Omega}{\alpha},\qquad
\widetilde n^2\rightarrow\frac{\widetilde n^2}{\alpha^2},
\qquad
d\rightarrow\alpha^2d.
\label{dScaling}
\end{equation}
Hence $\Omega^2/\widetilde n^2$ and $d\Omega^2$ remain unchanged, and
multiplication of $d$ by the positive real factor $\alpha^2$ cannot change
whether $\operatorname{Im}d=0$. Supplying $\omega_0$ therefore removes the
first ambiguity but does not remove this exceptional continuous ambiguity.

The singular case identified previously
\cite{tomoda2026simultaneous} is one member of this more general family.
Writing $\gamma=\widetilde P_{12}/(\widetilde n^2-1)$ gives
$d=(1+\gamma)/\widetilde n^2$. Hence
$\widetilde P_{12}=1-\widetilde n^2$ corresponds to $d=0$; in
Eq.~\eqref{exceptionalPfamily}, the original solution is then the
$\beta=0$ member. For $\beta>0$, $G_\beta(\Omega)$ has a zero when
$\Omega^2/4=1/\beta$, whereas for $\beta<0$ it has no real-frequency zero but
belongs to the same exactly nonunique family. The exceptional ambiguity is
therefore more general than the previously identified singular case.

A separate lower-dimensional cancellation occurs for
$\widetilde P_{12}=0$. Equation~\eqref{dDefinition} then gives
$d=1/\widetilde n^2$, so the numerator and denominator of
Eq.~\eqref{dimensionlessG} cancel and $G(\Omega)=1$. The echo then contains
no refractive-index information. This zero-photoelastic limit is distinct
from the real-$d$ photoelastic--strain ambiguity.

For the parameters used to generate the synthetic echoes in the main text,
$\widetilde n=1.47+1.95i$ and
$\widetilde P_{12}=1.98+10.74i$, giving
$d=-0.2721-0.3431i$. The synthetic example therefore lies away from the
exceptional $\operatorname{Im}d=0$ set. The reciprocal-conjugate alternative
is also nonpassive for these parameters, leaving a single passive material
solution once $\omega_0$ is supplied.

\end{document}